\documentstyle[12pt]{article}
\newcommand{\be}{\begin{equation}}
\newcommand{\ee}{\end{equation}}
\newcommand{\ba}{\begin{eqnarray}}
\newcommand{\ea}{\end{eqnarray}}

\newcommand{\D}{{\cal{D}}}
\newcommand{\Tr}{{\rm{Tr}}}
\newcommand{\pl}{\parallel}
\date{}
\renewcommand{\theequation}{\arabic{section}.\arabic{equation}}

\newcommand{\GeV}{\mbox{GeV}}

\newcommand{\grgl}{\:\hbox to -0.2pt{\lower2.5pt\hbox{$\sim$}\hss}
           {\raise3pt\hbox{$>$}}\:}
\newcommand{\klgl}{\:\hbox to -0.2pt{\lower2.5pt\hbox{$\sim$}\hss}
           {\raise3pt\hbox{$<$}}\:}

\begin{document}
\begin{titlepage}
\begin{flushright}
hep-th/9708051                    \\
\end{flushright}
\quad\\
\begin{center}
{\bf\LARGE Gluon Condensation}\\
\bigskip
{\bf\LARGE in Nonperturbative Flow Equations}\\
\vspace{1cm}
M. Reuter\\
\bigskip
DESY, Notkestra\ss e 85, D-22603 Hamburg \\
\vspace{.6cm}
C. Wetterich\\
\bigskip
Institut  f\"ur Theoretische Physik\\
Universit\"at Heidelberg\\
Philosophenweg 16, D-69120 Heidelberg\\
\vspace{3cm}
{\bf Abstract:}\\
\parbox[t]{\textwidth}{We employ nonperturbative flow equations for
an investigation of the effective action in Yang-Mills theories.
We compute the effective action $\Gamma[B]$ for constant color
magnetic fields $B$ and examine Savvidy's conjecture
of an unstable perturbative vacuum. Our results indicate that the
absolute minimum of $\Gamma[B]$ occurs for $B=0$. Gluon condensation
is described by a nonvanishing expectation value of the regularized
composite operator $F_{\mu\nu}F^{\mu\nu}$ which agrees
with phenomenological estimates.}
\end{center}\end{titlepage}
\newpage

\section{Introduction}
\setcounter{equation}{0}

A perturbative calculation of the effective action for constant
color-magnetic fields $B$ in Yang-Mills theories indicates that the
configuration
of lowest Euclidean action does not correspond to vanishing fields \cite
{sav},\cite{dit}.
This has led to many interesting speculations about the nature of the
QCD vacuum. Unfortunately, perturbation theory is clearly invalid in the
interesting region in field space. It breaks down both for vanishing
magnetic fields and for fields $B$ corresponding to the minimum of the
perturbatively calculated effective action $\Gamma[B]$. A nonperturbative
computation of $\Gamma[B]$ was lacking so far.
In this paper we make an attempt to evaluate $\Gamma[B]$
based on the non-perturbative method of the average action \cite{Av}.
Within our approximations we find that the minimum of $\Gamma[B]$
occurs for a vanishing magnetic field $B=0$, in contrast
to the perturbative result \cite{sav},
\cite{dit}. This strongly suggests that the
minimum of the effective action occurs for vanishing gauge
fields  $A_\mu=0$. This implies a vanishing expectation value
$<A_\mu>=0$ and Lorentz invariance of the QCD vacuum
is trivially realized\footnote{See ref. \cite{reuwe} for a discussion
of possible realizations of Lorentz symmetry for $<A_\mu>\not=0$.}.

The vanishing expectation value $<A_\mu>$ does by no means
indicate that the QCD vacuum is simple. Gluon condensation
phenomena may be described by nonvanishing expectation values
of composite operators such as $<F_{\mu\nu}F^{\mu\nu}>$.
We also discuss this issue in the context of the average action and
compute the expectation value of a suitably regularized operator.
Our result for the gluon condensate agrees
with phenomenological estimates. (The anomalous
trace of the energy momentum tensor induced by the condensate
$<F_{\mu\nu}F^{\mu\nu}>$ is found to be $-0.015\ {\rm GeV}^4$).
This may be
considered as a first analytical computation of this
condensate from ``first principles''. However, in view
of the large possible truncation
errors and the uncertainty in the precise correspondence
between the computed
operator and more phenomenological concepts of a gluon condensate it
seems premature to draw any quantitative conclusions. We rather
consider this work as an encouraging first step towards a
quantitatively more reliable computation of the gluon condensate.

The average
action $\Gamma_k$ is the effective Euclidean action for
averages of fields
which obtains by integrating out all quantum fluctuations with
(generalized)
momenta $q^2>k^2$. It can be viewed as the standard effective action
$\Gamma$ computed with an additional infrared cutoff $\sim k$ for all
fluctuations. In the limit $k\to 0$ the average action equals the
usual effective action, $\Gamma_0=\Gamma$. For $k>0$ no infrared
divergences
should appear in the computations and a lowering of $k$ allows to explore
the long-distance physics step by step. The $k$-dependence of the
average action is described by an exact nonperturbative evolution equation
\cite{Ev}.\footnote{For the relation to earlier versions of exact
renormalization group equations \cite{Rg} see ref. \cite{Mw}.}
The structure of this equation is close to a perturbative one-loop
equation
but it involves the full propagator and vertices instead of the classical
ones. For gauge theories it can be formulated in a way such that
$\Gamma_k[A]$
is a gauge-invariant functional of the gauge field $A$
\cite{reu}, \cite{cor}, \cite{top}.\footnote{See
ref. \cite{Gau}, \cite{Ell2} for alternative formulations.}

The flow equation for the pure non-abelian Yang-Mills theory has
already been solved
\cite{reu} with a very simple approximation - the average action
$\Gamma_k$
has been truncated with a minimal kinetic term $\sim
Z_kF_{\mu\nu}F^{\mu\nu}$. From the $k$-dependence of $Z_k$ the running of
the renormalized
gauge coupling $g(k)$ has been derived for arbitrary dimension $d$. Most
strikingly, this lowest-order estimate suggests that the wave function
renormalization $Z_k$ reaches zero for $k=k_\infty$ and turns negative
for $k<k_\infty$. Here the scale $k_\infty$ can be identified with the
confinement scale, i.e., the scale where the renormalized gauge coupling
$g^2(k)\sim Z_k^{-1}$ diverges. This was
interpreted as an indication for the instability of the perturbative
vacuum
with $A_\mu=0$, similar to the perturbative result \cite{sav}, \cite{dit}.
Indeed, negative $Z_k$ would imply that
the minimum of $\Gamma_k$ must occur for nonvanishing gauge field
for all $k<k_\infty$ \cite{reuwe}.  It is obvious, however, that
a truncated
effective action $\sim Z_kB^2$ is insufficient to describe
such phenomena. One expects positive $\Gamma_k$ for large $B$ and this
cannot be accommodated with a negative $Z_k$ in the ``lowest
order truncation''.

In this paper we enlarge the ``space of actions''
by considering for $\Gamma_k$ an arbitrary function of constant magnetic
fields, i.e., $\Gamma_k\sim W_k(\frac{1}{2}B^2)$. The flow equation
describes how the function $W_k$ changes its shape, starting from a linear
dependence $W_k=\frac{1}{2}Z_kB^2$ for large $k$.
The approximations (truncations) employed for the computation and
solution of the flow equation are summarized at the end of sect. 2.
In particular, we solve numerically the flow equations in a
truncation where $W_k$ is approximated by a polynomial of order
$B^6$. We find that in contrast to the perturbative computation
of ref. \cite{sav} the minimum of $W_k$ remains at $B=0$ for all $k$.
The vanishing of $Z_k$ for $k_\infty>0$ turns out to be an artifact
of the lowest order truncation. In the truncation of order $B^6$
the gauge coupling grows large but remains finite. In this approximation
the ground state value of the gauge field $A_\mu$ vanishes, in contrast
to speculations in ref. \cite{reuwe}.

For the description of gluon condensation in terms of
the expectation value of a suitably regularized gauge-invariant
composite operator $F_{\mu\nu}F^{\mu\nu}$ we use
exact flow equations including
composite operators \cite{Ell}.
Following this approach we enlarge in sect. 4 our truncation
by studying a general function $V_k(\frac{1}{2}B^2,\chi)$, where
$\chi$ is a composite field related to the operator
$F_{\mu\nu}F^{\mu\nu}$. We find that $\chi$ develops
indeed an expectation value which is stable for $k\to0$. This
implies a nonvanishing gluon condensate.

In the following we shall consider the flow equation for
the pure $SU(N)$
Yang-Mills theory.
The general formalism of our formulation with ghost fields and
background fields $\bar A$ is briefly reviewed in appendix A,
to which the reader could turn at this point. There
we write down the corresponding modified Slavnov-Taylor identities
\cite{Ell2} and a further identity which describes the dependence of
the effective action on the background field $\bar A_\mu$ \cite{wet},
\cite{rwe}.
We also give in this appendix the full exact flow equation
including the ghost sector. Since we are interested in the ground state
and there exist always classical solutions with vanishing ghost fields,
we concentrate here on the zero ghost sector. Then the effective action
reduces to a functional $\Gamma_k[A,\bar A]$ depending only on the
gauge field $A$ and the background field $\bar A$. It is invariant
under simultaneous gauge transformations of both fields. With a suitable
truncation for the ghost dependence of the full effective action (cf.
appendix A) one obtains the following renormalization group
equation for the scale
dependence of $\Gamma_k[A,\bar A]\ (t\equiv ln\ k)$:
\ba\label{1.1}
\frac{\partial}{\partial t}\Gamma_k[A,\bar
A]&=&\frac{1}{2}\Tr_{xcL}\left[\left(\Gamma^{(2)}_k[A,\bar
A]+R_k(\Gamma^{(2)}_k[\bar A,\bar A])\right)^{-1}\frac{\partial}{\partial
t} R_k(\Gamma^{(2)}_k[\bar A,\bar A])\right]\nonumber\\
&&-\Tr_{xc}\left[(-D^\mu[A]D_\mu[\bar A]+R_k(-D^2[\bar
A]))^{-1}\frac{\partial}{\partial t}R_k(-D^2[\bar A])\right]
\ea
Here the first trace on the r.h.s. arises from the fluctuations of the
gauge field. It involves on integration over space-time (``$x$'') as well
as a summation over color (``$c$'') and Lorentz (``$L$'') indices. The
``color'' trace is in the adjoint representation. The second trace is due
to the Faddeev-Popov ghosts and is over space-time and  color indices
only.  The precise form of the  infrared cutoff is described by the
function $R_k$. It is convenient to choose a smooth cutoff which vanishes
for large covariant momenta
\be\label{1.2}
R_k(u)=u[\exp({\cal Z}_k^{-1} u/k^2)-1]^{-1}
\ee
For the purpose of comparison we will also
consider in this paper a second choice for the cutoff, namely simply
a constant
\be\label{1.3}
R_k={\cal Z}_kk^2\ee
In eqs. (\ref{1.2}), (\ref{1.3})
the wave function renormalization
${\cal Z}_k$ could be a matrix in the space of fields which may even
depend on $\bar A$. (In ref. \cite{reu} we used ${\cal Z}_k\equiv 1$ for
the ghosts and a $k$-dependent constant ${\cal Z}_k\equiv Z_{F,k}$ for
all modes of the gauge field.) Eq. (\ref{1.1}) is a functional differential
equation where
$\Gamma_k^{(2)}[A,\bar A]$ denotes the matrix of second
functional derivatives of $\Gamma_k$ with respect to $A$, with the
background field $\bar A$ kept fixed. The modes  of the gauge field are
declared ``high-frequency modes''
or ``low-frequency modes'' depending on whether  their eigenvalues with
respect to the operator  $\Gamma_k^{(2)}[\bar A,\bar
A]\equiv\Gamma_k^{(2)}[A,\bar A]\bigm\vert_{A=\bar A}$ are larger or
smaller than $k^2$, respectively. Therefore it is this operator which
appears in the argument of $R_k$\footnote{In \cite{reu} we used the
classical $S^{(2)}[\bar A,\bar A]$ rather than $\Gamma_k^{(2)}[\bar A,\bar
A]$ for this purpose. While preserving all the general properties of
$\Gamma_k$, the new flow equation is much easier to handle from a
technical point of view.  This change also entails the different
positioning of ${\cal Z}_k$ in (\ref{1.2}) relative to the one in
\cite{reu}: for the simple truncation used there one has
$\Gamma_k^{(2)}=Z_{F,k}S^{(2)}$ for part of the modes. We only will
use the definition (\ref{1.2}) if for high momenta $q^2\to\infty$
one has $u\to\infty$. We
should mention, however, that for negative eigenvalues of $\Gamma_k^{(2)}
[\bar A,\bar A]$ the vanishing of $R_k$ for $k\to0$ is guaranteed only if
the
ratio $u/k^2$ in (\ref{1.2}) remains finite. This problem concerns
mainly the approach to convexity of the effective action for $k\to0$
and is of no relevance for the present work.}
and in this sense $R_k$ acts as an effective infrared cutoff by
suppressing the low frequency modes.

In this paper we aim for an approximate solution of the flow equation
(\ref{1.1}) by a truncation of the most general form of
$\Gamma_k[A,\bar A]$. We derive in sect. 2 the evolution equation
for the effective action for constant color magnetic fields $B$.
This equation is further approximated in sect. 3 by using a
polynomial ansatz for $\Gamma_k(B)$ of order $B^6$ and solving it
numerically. In sect. 4 we enlarge the truncation by introducing
a field for the composite operator $F_{\mu\nu}F^{\mu\nu}$.
This will permit the investigation of the flow of the expectation
value of $F_{\mu\nu}F^{\mu\nu}$ and an extraction of the gluon
condensate for $k\to0$. Sect. 5 finally contains our conclusions
and a brief discussion of further possible developments. Since
some of the necessary computations are technically involved,
we display a lot of this material in various appendices.

\section{Evolution equation of the
effective action for constant color-magnetic fields}
\setcounter{equation}{0}
In order to find nonperturbative approximative
solutions of (\ref{1.1}) we
employ the following ansatz for $\Gamma_k$:
\be\label{2.1}
\Gamma_k[A,\bar A]=\int d^dx\
W_k(\frac{1}{4}F^z_{\mu\nu}(x)F^{\mu\nu}_z(x))+
\frac{1}{2\alpha_k}\int d^dx\sum_z(D_\mu[\bar A](A^\mu-\bar A^\mu))^2_z\ee
Here $W_k$ is an arbitrary function of the invariant $\frac{1}{4}F^2$ with
$F^z_{\mu\nu}$ the field strength of the gauge field $A$. The
$k$-dependence
of $W_k$ will be determined by inserting (\ref{2.1}) into the evolution
equation (\ref{1.1}). If we think of $W_k(\theta)$,
\be\label{2.1a}\theta\equiv\frac{1}{4}F_{\mu\nu}^zF^{\mu\nu}_z,\ee
as a power series
in $\theta$ our ansatz contains at this point still
invariants of arbitrarily high
canonical dimension. Since all invariants which occur are of the form
$\theta^l$, only the dimensions $4l,l=1,2,...,$  actually occur
in this truncation. Also, for
a fixed dimension $4l$ the truncation (\ref{2.1}) does not contain
a complete basis of operators. Nevertheless one may hope that (\ref{2.1})
gives a qualitatively correct picture of the
effective action for constant color magnetic fields
in the regime where the
renormalization group evolution has already drastically modified the
classical Lagrangian  $\frac{1}{4}F^2$. We remark that effective actions
which depend on $\theta$ only play also a central role in the leading-log
models \cite{adl} of QCD.

The second term on the r.h.s. of (\ref{2.1}) is
a standard background gauge-fixing term \cite{abb}, \cite{reu} with a
$k$-dependent gauge-fixing parameter $\alpha_k$. In addition to the
$k$-dependence of the function $W_k(\theta)$ we should, in principle, also
compute the $k$-dependence  of $\alpha_k$. We will omit this here since
the general identities of appendix A imply that $\alpha$ is
independent
of $k$ in a first approximation. Furthermore, we will see that
within the truncation
(\ref{2.1}) and for a suitable choice of the infrared cutoff $R_k$ the
evolution equation for $W_k$ becomes independent of $\alpha_k$.

We should mention at this place that a truncation is actually not
completely defined
by the terms retained but rather by specifying which invariants in the
most general form of $\Gamma_k$ are omitted. One may parametrize a general
$\Gamma_k$ by infinitely many couplings multiplying the infinitely many
possible invariants which can be formed from the gauge fields consistent
with the symmetries. In the corresponding infinite dimensional space
a truncation is a projection on a subspace which is defined by setting all
but
the specified couplings to zero. (In our case the subspace remains
infinite dimensional.) In practice, we will choose a particular test
configuration $A_\mu$ corresponding to a constant magnetic field. The  
truncation should then be
understood in the sense that we use a basis for the invariants where all
invariants except those used in (\ref{2.1}) vanish for the test
configuration. By putting the coefficients of all invariants which
vanish for the test configuration to zero the truncation is uniquely
defined.
A computation of $W_k$ therefore amounts to a computation of the
$k$-dependent
effective action for a (particular) constant magnetic field.

As a particularly convenient test field we choose
a covariantly constant color-magnetic field \cite{dit} with a vector
potential of the form
\be\label{2.7}
A^z_\mu(x)=n^z{\sf A}_\mu(x)\ee
Here $n^z$ is a constant unit vector in color space $(n^zn_z=1)$, and
${\sf A}_\mu(x)$ is any ``abelian'' gauge field whose field strength
\be\label{2.8}
{\sf F}_{\mu\nu}=\partial_\mu{\sf A}_\nu-\partial_\nu{\sf A}_\mu=B\epsilon
^\perp_{\mu\nu}=const\ee
corresponds to a constant magnetic field $B$ along the 3-direction, say.
(We
define $\epsilon^\perp_{12}=-\epsilon^\perp_{21}=1$, with all other
components
vanishing.) Hence we have $\theta=\frac{1}{4}F^z_{\mu\nu}F^{\mu\nu}_z=
\frac{1}{2}B^2$.

In summary, the solution of the evolution equation will produce the
effective action for constant color magnetic fields of the type
(\ref{2.7}, \ref{2.8}), where an additional infrared cutoff $k$ is
present. One is finally interested in the limit $k\to 0$. For the
computation of this effective action we make approximations which amount
to the following truncations of the effective action:
\begin{enumerate}

\item The ghost sector is approximated by
its classical form, as discussed in appendix A.

\item The remaining gauge field dependence
of $\Gamma_k$ is approximated by (\ref{2.1}).
\item The resulting flow equation which will be derived in this
section is a nonlinear partial differential equation
for a function of two variables, $W_k(\theta)\equiv W(k,\theta)$.
We solve numerically in sect. 3
only an approximate version of this differential
equation where $W_k(\theta)$ is truncated to a polynomial in $\theta$ of
degree three.
\end{enumerate}

We do not expect these approximations to yield a quantitatively precise
result in the range of $k$ where the gauge coupling is large.
Nevertheless, it remains an interesting question if the speculated effect
of an unstable perturbative vacuum persists
in this picture. Our findings indicate, in
contrast to perturbation theory \cite {sav}, \cite{dit} ,
that the minimum of the effective action occurs for $A_\mu=0$.  We
emphasize that despite the approximations made our approach goes far
beyond the perturbative calculation of refs. \cite{sav},\cite{dit}.
We conclude that there is no reason to believe that the
configuration $B=0$ (or $A_\mu=0$) is unstable.

The evolution equation for $W_k(\theta)$ is computed
for arbitrary dimension $d$ in appendix B. One finds
\ba\label{3.4}
\frac{\partial}{\partial t}W_k(\frac{1}{2}B^2)&=&\frac{1}{2}\Omega^{-1}
\Tr_{xcL}[H(W_k'{\cal D}_T)]\nonumber\\
&&+\frac{1}{2}\Omega^{-1}
\Tr_{xc}[\tilde H(-\alpha_k^{-1}D^2)-H(-W_k' D^2)]
-\Omega^{-1}\Tr_{xc}[H_G(-D^2)]\nonumber\\
&&+v_d\left(\frac{1}{W_k'+B^2W_k''}-\frac{1}{W_k'}\right)
\left(\frac{1}{W_k'}
\right)^{\frac{d}{2}-1}\int^\infty_0dx x^{\frac{d}{2}-1}H(x)\nonumber\\
&&-v_d\int^\infty_0dx x^{\frac{d}{2}-1}(\tilde
H\left(\frac{x}{\alpha_k}\right)
-H\left(\frac{x}{\alpha_k}\right))\ea
with $W_k'\equiv(\partial W_k/\partial\theta)(\frac{1}{2}B^2)$,
etc., $\Omega=\int d^dx$, and
\be\label{3.5}
v_d^{-1}\equiv
2^{d+1}\pi^{\frac{d}{2}}\Gamma\left(\frac{d}{2}\right)\ee
Here we have introduced the convenient abbreviation
(for ${\cal Z}_k=Z_k$)
\ba\label{3.3}
H(u)&\equiv&(u+R_k(u))^{-1}\frac{\partial}{\partial t}R_k(u)\nonumber\\
&=&\left\lbrace
\begin{array}{lll}
(2+\frac{d}{dt}\ln
Z_k)\frac{u}{Z_kk^2}\left[\exp\left(\frac{u}{Z_kk^2}\right)-1\right]^{-1}&
{\rm for}& (1.2)\\
(2+\frac{d}{dt}\ln Z_k)Z_kk^2[u+Z_kk^2]^{-1}&{\rm for}&(1.3)\end{array}
\right.
\ea
In the second, ghost-type, trace the function $H_G(u)$ is defined
similarly,
but with a different factor ${\cal Z}_k=1$. The last trace accounts for a
possible difference between $\tilde Z_k$ and $Z_k$ in (\ref{3.1}).
Here $\tilde H$ is obtained from $H$ by replacing $Z_k$ by $\tilde Z_k$.

The eigenvalues of the operator
\[{\cal D}_T
\equiv-D^2+2i\bar gF\]
are known explicitly \cite{dit}.
They are parametrized by a $(d-2)$-dimensional momentum $q^\mu$ which
``lives'' in the space orthogonal to the 1-2 plane,
and a discrete quantum
number $n=0,1,2,...$ which labels the Landau levels. The spectral sum
for the function $\hat H(x)\equiv H(W_k'x)$ reads\footnote{The
 momentum integration is absent for $d=2$.}
\ba\label{3.6}
\Omega^{-1}\Tr_{xcL}[\hat H({\cal D}_T)]&=&\sum^{N^2-1}_{l=1}\frac{\bar
g|\nu_l|B}{2\pi}\sum^\infty_{n=0}\int
\frac{d^{d-2}q}{(2\pi)^{d-2}}\nonumber\\
&&\cdot\left\lbrace (d-2)\hat H(q^2+(2n+1)\bar
g|\nu_l|B)\right.\nonumber\\
&&+\hat H (q^2+(2n+3)\bar g|\nu_l|B)\nonumber\\
&&\left.+\hat H (q^2+(2n-1)\bar g|\nu_l|B)\right\rbrace\ea
Here $\nu_l,l=1,..., N^2-1$ are the eigenvalues of the matrix $n^zT_z$ in
the adjoint representation. We note that for $n=0$ and $q^2$ sufficiently
small the eigenvalue $q^2-\bar g|\nu_l|B$ in the third term on the r.h.s.
of
(\ref{3.6}) can become negative. This instability \cite{sav},
\cite{nol} causes severe problems
if one tries to compute the standard one-loop effective action in
the background of a covariantly constant magnetic field \cite{ksw}.
In our approach
this problem is cured by the presence of an IR regulator.
Eq. (\ref{3.6}) can be rewritten $(d>2)$ as
\ba\label{3.7}
&&\Omega^{-1}\Tr_{xcL}[\hat H({\cal D}_T)]=\frac{v_{d-2}}{\pi}\sum^{N^2-1}
_{l=1}\bar g|\nu_l|B\int^\infty_0dx x^{\frac{d}{2}-2}\nonumber\\
&&\cdot \left\lbrace d\sum^\infty_{n=0}\hat H(x+(2n+1)\bar g|\nu_l|B)+\hat
H(x-\bar g|\nu_l|B)-\hat H(x+\bar g|\nu_l|B)\right\rbrace\ea
but it cannot be simplified any further in closed form. The other traces
in (\ref{3.4})
are given by
\be\label{3.8}
\Omega^{-1}\Tr_{xc}[\hat H(- D^2)]=\frac{v_{d-2}}{\pi}\sum^{N^2-1}
_{l=1}\bar g|\nu_l|B\int^\infty_0dx x^{\frac{d}{2}-2}
\sum^\infty_{n=0}\hat H(x+(2n+1)\bar g|\nu_l|B)\ee
and similarly for $\tilde H$ and $H_G$.

In this paper we use two different methods in order to (approximately)
compute
the spectral sums (\ref{3.7}) and (\ref{3.8}). In appendix F we shall
represent
them as Schwinger proper-time integrals \cite{direu}. This method leads to
compact
integral representations which are valid for all values of $B$, but
it has the disadvantage that it works only for
the cutoff function $R_k(x)$ defined by (\ref{1.6}) which leads,
as
we shall see, to ultraviolet problems.

The second method consists
of expanding the r.h.s. of (\ref{3.7}) in powers
of $B$. It is applicable
if $\bar g B\ll k^2$.  It works for any function $R_k(x)$
such that ultraviolet problems can easily be avoided. The
condition $\bar g B\ll k^2$ guarantees that we may express the sum
over $n$ by an Euler-McLaurin series, and that the terms
become small rapidly. We will concentrate mainly on the
second approach. In this
manner (\ref{3.7}) turns into
\ba\label{3.9}
&&\Omega^{-1}\Tr_{xcL}[H(W_k'(\frac{1}{2}B^2){\cal D}_T)]\nonumber\\
&&=(N^2-1)\frac{dv_{d-2}}{2\pi}[W_k'(\frac{1}{2}B^2)]^{-\frac{d}{2}}
\int^\infty_0 dx\int^\infty_0dy x^{\frac{d}{2}-2}H(x+y)\\
&&+\frac{v_{d-2}}{\pi}\sum^\infty_{m=1}C^d_m\left(\sum^{N^2-1}_{l=1}
\nu^{2m}_l\right)(\bar g
B)^{2m}[W_k'(\frac{1}{2}B^2)]^{2m-\frac{d}{2}}\int^\infty_0dx\
x^{\frac{d}{2}
-2}H^{(2m-1)}(x)\nonumber\ea
with
\be\label{3.10}
C^d_m=\frac{d}{(2m)!}(2^{2m-1}-1)B_{2m}-\frac{2}{(2m-1)!}\ee
Here $B_{2m}$ are the Bernoulli numbers. In a second step one has to
expand
the $B^2$-dependence of $W_k'$. Note that only even powers of $B$ occur in
this
expansion. The group-theoretical factors $\sum^{N^2-1}_{l=1}\nu^{2m}_l$
are discussed in appendices C and D. In particular for
$SU(2)$ these factors equal 2 for all values of $m$.

If we use the Euler-McLaurin series (\ref{3.9}) and a similar expansion for the
trace (\ref{3.8}) in eq. (\ref{3.4}), we find for $SU(N)$:
\ba\label{3.11}
&&\frac{\partial}{\partial t}W_k(\theta)=\frac{v_{d-2}}{2\pi}(2-\eta)k^d
\left\lbrace\frac{d-1}{2}(N^2-1)r^d_2\left(\frac{W_k'}
{Z_k}\right)^{-\frac{d}{2}}\right.\nonumber\\
&&\left.-\sum^\infty_{m=1}\tau_m(C^d_m-E_m)r^{d,m}_0\left(\frac
{2\bar
g^2\theta}{k^4}\right)^m\left(\frac{W_k'}{Z_k}\right)^{2m-\frac{d}{2}}
\right\rbrace\nonumber\\
&&+\frac{v_{d-2}}{2\pi}(2-\tilde \eta)(\tilde
Z_k\alpha_k)^{\frac{d}{2}}k^d\left\lbrace\frac{1}{2}(N^2-1)r^d_2\right.
\left.-\sum^\infty_{m=1}\tau_m E_mr^{d,m}_0\left(\frac
{2\bar g^2\theta}{k^4}\right)^m(\tilde Z_k\alpha_k)^{-2m}
\right\rbrace\nonumber\\
&&-\frac{v_{d-2}}{\pi}k^d\left\lbrace(N^2-1)r^d_2-2\sum^\infty_{m=1}
\tau_m E_mr^{d,m}_0\left(\frac
{2\bar g^2\theta}{k^4}\right)^m
\right\rbrace\nonumber\\
&&+v_d(2-\eta)r^d_1k^d\left(\frac{Z_k}{W_k'+2\theta
W_k''}-\frac{Z_k}{W_k'}\right)\left(\frac{W_k'}{Z_k}
\right)^{1-\frac{d}{2}}
+\ {\rm const.}\ea
with $\tau_m$ defined in appendix D and the
constants $E_m$ given by
\be\label{3.12}
E_m=\frac{1}{(2m)!}(2^{2m-1}-1)B_{2m}.\ee
The dimensionless integrals
\ba\label{3.13}
r^{d,m}_0&=&-\frac{1}{2-\eta}(Z_kk^2)^{2m-\frac{d}{2}}
\int^\infty_0dxx^{\frac{d}{2}-2}H^{(2m-1)}(x)\nonumber\\
&=&-\int^\infty_0dx\ x^{\frac{d}{2}-2}\left(\frac{d}{dx}\right)^{2m-1}
\frac{x}{e^x-1}\nonumber\\
r^{d}_1&=&\frac{1}{2-\eta}(Z_kk^2)^{-\frac{d}{2}}
\int^\infty_0dxx^{\frac{d}{2}-1}H(x)=\int^\infty_0dx\
\frac{x^{\frac{d}{2}}}{e^x-1}\nonumber\\
r^d_2&=&\frac{1}{2-\eta}(Z_kk^2)^{-\frac{d}{2}}\int^\infty_0dx
\int^\infty_0dyx^{\frac{d}{2}-2}H(x+y)\nonumber\\
&=&
\int^\infty_0dx\int^\infty_0dy\frac{x^{\frac{d}{2}-2}(x+y)}{\exp(x+y)-1}
\ea
occur as a consequence of (\ref{3.3}). The second equality in eqs.
(\ref{3.13}) uses  the exponential cutoff (\ref{1.2}).
For this choice\footnote{In contrast,
the quantities $r^4_1$ and $r^4_2$ are not well defined for the choice
(\ref{1.6}). The ultraviolet divergence indicates an incomplete ``thinning
out'' of the high momentum degrees of freedom for a simple mass like
infrared cutoff. Even though eq. (\ref{3.11})
was derived by choosing a specific background field, this
evolution equation  does not depend on the background we used for the
calculation. By employing derivative expansion techniques \cite{zuk},
\cite{mgs} it should also be possible to derive (\ref{3.11}) without ever
specifying a  background. In the case at hand the method presented here
is by far simpler, however.}
we note for  later use that in 4 dimensions
\be\label{3.14}
r^{4,m}_0=B_{2m-2},\ r^4_1=r^4_2=2\zeta(3)\ee
where $\zeta$ denotes the Riemann zeta function.
The evolution equation (\ref{3.11}) is the central result of this
section. It constitutes a partial differential equation for a
function of two variables, $W(\theta,k)$.

In eq. (\ref{3.11}) we have introduced the anomalous dimensions
\be\label{3.15}
\eta=-\frac{d}{dt}\ln Z_k,\quad\tilde\eta=-\frac{d}{dt}\ln \tilde Z_k\ee
A convenient choice for the wave function renormalization constants
used in the infrared cutoff $R_k$ is
\ba\label{3.16}
Z_k&=&\left\lbrace\begin{array}{ccc}
W_k'(0)&\ {\rm for}&\ k>k_{np}\\
W_{k_{np}}'(0)&\ {\rm for}&\ k<k_{np}\end{array}\right.\nonumber\\
\tilde Z_k&=&\frac{1}{\alpha_k}\ea
where $k_{np}$ is a typical momentum scale which characterizes the
transition from the perturbative to the nonperturbative regime. We
account for the possibility that $W_k'(0)$ may turn negative for
$k$ smaller than a ``confinement scale'' $k_\infty$, whereas $Z_k$ must
always be strictly positive.  More precisely, if $|d\ln W_k'(0)/dt|$
becomes of order unity for small scales $k$, we
choose $k_{np}$ to be the scale where
$|\eta(k_{np})|=1.5$. For $k<k_{np}$ all couplings run fast anyhow,
and an improvement of the scaling properties of $R_k$ by the introduction
of a $k$-dependent wave function renormalization seems not necessary.
The choice $\tilde Z_k=\alpha_k^{-1}$ guarantees
that the infrared cutoff acts on the longitudinal modes in the same way
as on the transversal modes. It implies that the flow equation for
$\frac{\partial}{\partial t} W_k(\theta)$ becomes
independent\footnote{A $\theta$-independent constant in $W_k$
is irrelevant.} of $\alpha_k$ and therefore independent
of the ``gauge fixing'' in our
truncation! In this paper we can therefore
neglect the running\footnote{This can be inferred from a first-order
approximation to the solution of the general identities which govern
the dependence of $\Gamma_k[A,\bar A]$ on the background field
$\bar A$ \cite{wet}, \cite{rwe}.}
of $\alpha_k$, and one has $\tilde\eta=0$.

It is convenient to express the flow equation in terms of
renormalized
dimensionless quantities
\ba\label{3.17}
g^2&=&k^{d-4}Z^{-1}_k\bar g^2\nonumber\\
\vartheta&=&g^2k^{-d}Z_k\theta\nonumber\\
w_k(\vartheta)&=&g^2k^{-d}W_k(\theta)\ea
Switching to a notation where dots denote derivatives with respect
to $\vartheta$ instead of $\theta$ and with $\partial/\partial t$  now
taken at fixed $\vartheta$ one obtains
\ba\label{3.18}
&&\frac{\partial}{\partial t}w_k(\vartheta)=-(4-\eta)w_k(\vartheta)+
4\vartheta\dot w_k(\vartheta)\nonumber\\
&&+(2-\eta)v_dg^2(\dot w_k(\vartheta))^{-\frac{d}{2}}\left\lbrace
\frac{(d-1)(d-2)}{2}(N^2-1)r^d_2\right.\nonumber\\
&&\left.-\frac{2r^d_1\vartheta\ddot w_k(\vartheta)}{\dot w_k(\vartheta)+
2\vartheta\ddot w_k(\vartheta)}-(d-2)\sum^\infty_{m=1}\tau_m
(C^d_m-E_m)r^{d,m}_0(2\vartheta\dot
w_k^2(\vartheta))^m\right\rbrace\nonumber\\
&&+2(d-2)v_dg^2\sum^\infty_{m=1}\tau_mE_mr^{d,m}_0(2\vartheta)^m+{\rm
const}
\ea
This nonlinear partial differential equation for the function
$w(\vartheta,t)$ does not show an explicit $t$- (or $k$)-dependence
of the right-hand side any more.

One needs in
addition the running of the renormalized gauge coupling $g$ and
the anomalous dimension $\eta$,
which are related by
\be\label{3.19}
\beta_{g^2}=\frac{\partial g^2}{\partial t}=(d-4+\eta)g^2\ee
For $k>k_{np}$ we have by definition $\dot w_k(0)=1$ and $\eta$ can be
determined by
\ba\label{3.20}
&&\frac{\partial}{\partial t}\dot
w_k(0)=0=\eta-2(d-2)v_dr^{d,1}_0\tau_1g^2\left((2-\eta)C^d_1-(4-\eta)E_1
\right)\nonumber\\
&&-(2-\eta)v_dg^2\left(2r^d_1+\frac{d(d-1)(d-2)}{4}(N^2-1)r^d_2\right)
\ddot w_k(0)\ea
With $\tau_1=N,C^d_1=\frac{d}{12}-2,E_1=\frac{1}{12}$ this yields
\ba\label{3.21}
\eta=&-&\left(\frac{N}{3}v_d(d-2)(26-d)r^{d,1}_0g^2-2h_dg^2w_2\right)
\nonumber\\
&&\left(1-\frac{N}{6}v_d(d-2)(25-d)r^{d,1}_0g^2+h_dg^2w_2\right)^{-1}
\ea
where
\be\label{3.22}
h_d=v_d\left(2r^d_1+\frac{d(d-1)(d-2)}{4}(N^2-1)r^d_2\right)\ee
\be\label{3.23}
w_2=\ddot w_k(0)\ee
For general $d$ the constants $r^{d,1}_0,r^d_1$ and $r^d_2$ depend
on the precise choice of the infrared cutoff except
for $d=4$ where $r^{4,1}_0=1$ is cutoff independent. The running of the
renormalized gauge coupling $g$ is now fully determined by eq.
(\ref{3.19}). It
depends on the additional coupling $w_2$ (\ref{3.23}) which will be
discussed
in more detail in the next section\footnote{For $w_2=0$ we recover the
result
of ref. \cite{reu} except for the factor $(25-d)$ in the denominator
in (\ref{3.21}) which was $(24-d)$ previously. This difference is due to
a slightly different choice of the $Z$-factors in the infrared cutoff.}.

Specifying the initial value $g^2(\Lambda)$ and the function
$w_\Lambda(\vartheta)$ at some high momentum scale $\Lambda$ the form of
$w_k(\vartheta)$ and $g^2(k)$ are completely determined by the flow
equation (\ref{3.18}). Solving for $k\to 0$ the function
$w_0(\vartheta)$ specifies the effective action in our truncation. If
necessary, one has to replace for $k<k_{np}$ eq. (\ref{3.21}) by
$\eta=0,g^2(k<k_{np})=g^2(k_{np})$.

Before closing this section, we briefly comment on the range of
convergence
of the Euler-McLaurin series in our case. For large $m$ one has
\be\label{3.24}
\lim_{m\to\infty}C^d_m=d\lim_{m\to\infty}E_m\sim\pi^{-2m}\ee
For $N=2$ and $d=4$ we find (cf. (\ref{3.14})) that the coefficients
of $(2\vartheta)^m$ in eq. (\ref{3.18}) diverge $\sim\pi^{-2m}B_{2m-2}
\sim\pi^{-4m}2^{-2m}(2m-4)!$. For small nonvanishing $\vartheta$ the
first terms of the series have a very rapid apparent convergence,
but the series finally diverges due to the factorial growth $\sim
(2m-4)!$. We can therefore safely use this series only for the derivatives
$w^{(n)}(\vartheta=0)$ with finite $n$ where convergence problems are
absent since only a finite number of terms in the sum contributes.
The situation is probably similar for $N>2$ and/or $d\not=4$ as well
as for many other choices of the infrared cutoff. In contrast, the
original sums over $n$ in eqs. (\ref{3.7}), (\ref{3.8}) always converge
since for $B>0$ the contributions from sufficiently high values of $n$
are exponentially suppressed. An explicit evaluation of these sums
is possible for the simplified masslike IR-cutoff (\ref{1.6}). This
is described in appendix F where we will also see the reason for the
ultraviolet divergence of $r_1^4$ and $r^4_2$ for this particular cutoff.

\section{Polynomial truncations}
\setcounter{equation}{0}
In this section we concentrate on $d=4$ with the exponential
infrared cutoff (1.2).
One could solve the flow equation (\ref{3.18}) numerically. Instead,
we further simplify here the truncation in
order to get a first idea of the physical contents of
(\ref{3.11}).
We include in $w_k(\vartheta)$ only terms which are at most quadratic
in $\vartheta$:
\ba\label{4.1}
w_k(\vartheta)&=&w_0(k)+w_1(k)\vartheta+\frac{1}{2}w_2(k)\vartheta^2
+\frac{1}{6}w_3(k)\vartheta^3\nonumber\\
&&w_j(k)\equiv \left(\frac{d}{d\vartheta}\right)^jw_{k|\vartheta=0}\ea
Thus the truncation for $\Gamma_k$ is parametrized
(up to an irrelevant constant) by three couplings:
\ba\label{4.2}
\Gamma_k[A,A]&=&\int d^dx\left\lbrace
\frac{1}{4}\frac{\bar g^2w_1(k)}{g^2(k)}
F^z_{\mu\nu}F^{\mu\nu}_z\right.
+\frac{1}{32}\frac{\bar g^4 w_2(k)}
{g^2(k)k^4}\left(F^z_{\mu\nu}F^{\mu\nu}_z\right)^2\nonumber\\
&+&\left.\frac{1}{384}\frac{\bar g^6w_3(k)}{g^2(k)k^8}
\left(F^z_{\mu\nu}F^{\mu\nu}_z\right)^3\right\rbrace
\ea
The short distance or
``classical'' theory is specified for $k=\Lambda$ by $(Z_\Lambda=1)$
\be\label{4.3}
g^2(\Lambda)=\bar g^2,\quad w_1(\Lambda)=1,\quad
w_2(\Lambda)=w_3(\Lambda)=0\ee
For $k>k_{np}$ it follows from the definitions (\ref{3.16})
and (\ref{3.17}) that $w_1(k)=1$,
and $g^2(k)$ is determined by eqs. (\ref{3.19}), (\ref{3.21}). For
$k<k_{np}$
we use $g^2(k)=g^2(k_{np})$ instead and keep
\be\label{4.6}
w_1(k)\equiv\dot w_k(0)\ee
as the independent
running coupling constant
The flow equations for the partial derivatives
$w_j(k)\equiv w_k^{(j)}(\vartheta=0)$ follow  by differentiating eq.  
(\ref{3.18}) with respect to $\vartheta$.
For example, one has
\ba\label{4.7}
&&\frac{\partial}{\partial t}\dot w=\eta\dot w+4\vartheta\ddot
w-(2-\eta)v_dg^2\dot w^{-\frac{d}{2}}\cdot\nonumber\\
&&\left\lbrace (d-2)\sum^\infty_{m=1}
\tau_m(C^d_m-E_m)\right. r^{d,m}_0(2\vartheta\dot w^2)^{m-1}(2m\dot
w^2+(4m-d)\vartheta\dot w\ddot w)\nonumber\\
&&+r^d_1\left(\frac{2\ddot w+2\vartheta w^{(3)}}{\dot w+2\vartheta\ddot
w}-
\frac{2\vartheta\ddot w(3\ddot w+2\vartheta w^{(3)})}{(\dot
w+2\vartheta\ddot w
)^2}-d\frac{\vartheta(\ddot w)^2}{\dot w(\dot w+2\vartheta\ddot w)}\right)
\nonumber\\
&&\left.+\frac{1}{4}d(d-1)(d-2)(N^2-1)r^d_2\frac{\ddot w}{\dot w}
\right\rbrace\nonumber\\
&&+4(d-2)v_dg^2\sum^\infty_{m=1}m\tau_mE_mr^{d,m}_0(2\vartheta)^{m-1}
\ea
We observe that $w_0$ does not appear on the r.h.s. of the evolution
equations for $w_j, j\geq1$, and we omit in the following this irrelevant
constant.

For $k>k_{np}$ the evolution equation for the running gauge
coupling reads $(d=4)$
\be\label{4.8}
\frac{\partial g^2}{\partial t}=-\frac{g^4}{24\pi^2}\left(11N-
3H_4w_2\right)\left[1-\frac{g^2}{32\pi^2}
\left(7N-2H_4w_2\right)\right]^{-1}\ee
whereas for $k<k_{np}$ one uses the flow equation for $w_1$
\be\label{4.9}
\frac{\partial}{\partial
t}w_1=\frac{g^2(k_{np})}{8\pi^2}\left(\frac{11N}{3}
-H_4\frac{w_2}{w_1^3}\right)\ee
where $H_4=16\pi^2h_4=r^4_1+3(N^2-1)r^4_2=2(3N^2-2)\zeta(3)$
for the choice (\ref{1.2}). It is obvious that for $w_2<0$ the gauge
coupling $g^2(k)$ always increases (\ref{4.8}) until at $k=k_{np}$ the
anomalous dimension $|\eta|$ reaches 1.5. If $w_2$ remains negative for
$k<k_{np}$, the coupling $w_1$ decreases until it reaches zero at the
confinement scale $k_\infty>0$. (If we define $g^2(k)=g^2(k_{np})/w_1$,
this coupling diverges at the confinement scale.)
The issue is different for $w_2>0$: The coupling $w_1$ does not reach
zero
since for small enough $w_1$ the second term in eq. (\ref{4.9}) would
cancel
the first term. One therefore needs an estimate of $w_2(k)$.

Let us first consider the regime
$k>k_{np}$ where the evolution equation for $w_2$
reads
(with $E_2=-\frac{7}{720},\ C^4_2=-\frac{67}{180}$)
\ba\label{4.10}
&&\frac{\partial}{\partial
t}w_2=(4+\eta)w_2+\frac{g^2}{8\pi^2}\left\lbrace
\tau_2r_0^{4,2}\left(\frac{127}{45}-\frac{29}{20}\eta\right)\right.\\
&&\left.+(2-\eta)\left(5r^4_1+\frac{9}{2}(N^2-1)r^4_2\right)
w^2_2-(2-\eta)\left(r^4_1+\frac{3}{2}(N^2-1)r^4_2\right)w_3\right
\rbrace\nonumber\ea
We observe the appearance of the coupling $w_3$
whose evolution is given by
(for details see appendix F)
\ba\label{5.2}
&&\frac{\partial}{\partial
t}w_3=(8+\eta)w_3+\frac{g^2}{16\pi^2}\left\lbrace
-\frac{1}{30}\left(\frac{442}{315}-\frac{137}
{210}\eta\right)\tau_3\right.\\
&&\left.+\frac{87}{30}(2-\eta)\tau_2w_2-6\zeta(3)
(2-\eta)[(12N^2+23)w^3_2-
(9N^2+8)w_2w_3+N^2w_4]\right\rbrace\nonumber\ea
We neglect the term $\sim w_4$ which is consistent with our approximation.
In the perturbative region, where
$Ng^2/16\pi^2$ is small, it is easy to infer from (\ref{4.10})
that $w_2$ is of the order
$g^2$. In fact, we may neglect $\eta$ and
the terms $\sim w^2_2$ and $\sim w_3$ in the
curly bracket in (\ref{4.10}).
In lowest order one finds for the ratio $\frac{w_2}{g^2}$ an
infrared stable fixed point:
\be\label{4.11}
\frac{\partial}{\partial
t}\left(\frac{w_2}{g^2}\right)=4\frac{w_2}{g^2}+\frac{
127}{360\pi^2}\tau_2 r^{4,2}_0\ee
\be\label{4.12}
w_{2*}(k)=-\frac{127}{1440\pi^2}\tau_2r^{4,2}_0g^2(k)
=-\frac{127}{270}\frac{g^2}{16\pi^2}\ee
which is approached very rapidly. (The last equality in (\ref{4.12})
holds for $N=2$ and uses $r^{4,2}_0=\frac{1}{6}$ for the exponential
cutoff.) Similarly, one obtains the lowest order
fixed point
\be\label{5.3}
w_{3*}(k)=\frac{221\tau_3}{37800}\frac{g^2(k)}{16\pi^2}\ee
Actually, as we show in Appendix F,
all ratios $w_n/g^2$ reach perturbative fixed points
for $n\geq3$. This justifies the approximation
(\ref{4.1}) at least for small enough $g^2(k)$. (We observe $W_n\sim
g^{2(n-1)}k^{-4(n-1)}$.)

It is interesting to insert the value (\ref {4.12}) into the
$\beta$-function for $g^2$ eq. (\ref{4.8}). Expanding in
powers of $g^2$ one has
\ba\label{4.13}
&&\frac{\partial g^2}{\partial
t}=-\frac{22N}{3}\frac{g^4}{16\pi^2}
-\frac{77N^2}{3}\frac{g^6}{(16\pi^2)^2}
+(3N^2-2)\zeta(3)\frac{g^4}{4\pi^2}w_2\nonumber\\
&&=-\frac{22}{3}\frac{g^4N}{16\pi^2}-
\left(\frac{77}{3}+\frac{127}{45}\zeta
(3)\tau_2\left(1-\frac{2}{3N^2}\right)\right)\frac{g^6N^2}{(16\pi^2)^2}
\ea
We note that without the term $\sim w_2$ the coefficient $\sim g^6$
exceeds the perturbative two-loop coefficient
$-\frac{204}{9}\frac{N^2}{(16\pi^2)^2}$ only
by a little more than 10\%. It is
recomforting to find  the contribution from $w_2$ in
the same order of magnitude as this difference.  We
emphasize that a full computation of $\beta_{g^2}$ in order $g^6$ should
take additional invariants into account, as for example $(F\tilde F)^2$
or $(D_\mu F^{\mu\nu})^2$. We also observe that for $k=k_{np}$, i.e., for
$\eta=-\frac{3}{2}$ one has
approximately $Ng^2/16\pi^2=\frac{1}{7}$ so that the validity
of perturbation theory extends roughly to all $k>k_{np}$.

Let us finally consider the regime $k<k_{np}$. Our truncation permits,
in principle, a first-order type transition where the absolute
minimum of $w$ jumps from $\vartheta=0$ to a nonzero value
$\vartheta_0>0$. For
positive $w_3$ the polynomial (\ref{4.1}) is bounded from below for
$\vartheta\geq0$. For $w_1>0$ there is always a local minimum at
$\vartheta
=0$. Two additional extrema are present if
\be\label{5.4}
 w^2_2>2w_1w_3\ee
There is a minimum for positive $\vartheta_0$
\be\label{5.5}
\vartheta_0=\frac{-w_2+\sqrt{w^2_2-2w_1w_3}}{w_3}\ee
and, for $w_1>0$, a maximum at
\be\label{5.6}
\vartheta_{max}=\frac{-w_2-\sqrt{w_2^2-2w_1w_3}}{w_3}\ee
For $w_1<0$ the origin $\vartheta=0$ turns to a local
maximum. The critical set of couplings where the two minima
are of equal height $(w_k(\vartheta_0)=0)$ corresponds to
\be\label{5.7}
w_1=\frac{3}{8}\frac{w^2_2}{w_3}\ee
For $w_3$ smaller than the critical value, the absolute minimum
occurs at $\vartheta_0$. This would correspond to the
picture where
the perturbative vacuum is unstable, i.e.
$<A_\mu>\not=0$.

For $k>k_{np}$ where $w_1=1$ an inspection of the flow equations
(cf. (\ref{4.12}), (\ref{5.3})) shows that $2w_3$ remains larger
than $w^2_2$. There is therefore only one minimum at $\vartheta=0$.
If $w_2$ remains negative for $k<k_{np}$, the coupling $w_1$
would decrease towards zero. If furthermore $w_3$ stays positive,
the condition (\ref{5.7}) would then necessarily be met for some
scale $k>k_\infty$. On the other hand, if $w_2$ turns positive,
$w_1$ will also remain strictly positive, and the only minimum
occurs at $\vartheta=0$ for all values of $k$.

We have solved
numerically the system of flow equations for $k<k_{np}$
\ba\label{5.8}
&&\frac{\partial}{\partial
t}w_2=4w_2+\frac{g^2(k_{np})}{16\pi^2w_1^2}\Bigl
\lbrace\tau_2\left(\frac{29}{30}w^4_1-\frac{7}{270}w^2_1\right)
\nonumber\\
&&+4\zeta(3)[(9N^2+1)\frac{w^2_2}{w^2_1}-(3N^2-1)
\frac{w_3}{w_1}]\Bigr\rbrace\nonumber\\
&&\frac{\partial}{\partial
t}w_3=8w_3+\frac{g^2(k_{np})}{16\pi^2w^2_1}\left
\lbrace-\frac{\tau_3}{30}\left(\frac{137}{105}w^6_1+\frac{31}{315}
w^2_1\right)+\frac{87}{15}\tau_2w^3_1w_2\right.\nonumber\\
&&\left.-12\zeta(3)\left(N^2\frac{w_4}{w_1}-(9N^2+8)\frac{w_2w_3}{w_1^2}+
(12N^2+23)\frac{w_2^3}{w_1^3}\right)\right\rbrace\ea
together with eq. (\ref{4.9}) for $w_1$. (According to our
truncation, we put $w_4=0$.) For $N=3$ we
find that $w_3$ remains positive for
all $k$, whereas $w_2$ changes sign for $k\approx 0.5 k_{np}
\approx \Lambda_{\rm QCD}$. In
consequence, all three couplings run to fixed values as $k\to0$,
given by
\be\label{5.9}
w_{1*}=0.13,\quad w_{2*}=4\cdot 10^{-4},\quad w_{3*}=4\cdot10^{-6}\ee
At least within the $F^6$ truncation we find that the ground state
occurs for $A_\mu=0$, contradicting earlier speculations
and the estimates of too simple $F^2$ or $F^4$ truncations!
As an immediate consequence, the gauge coupling never diverges
and
\[\alpha_s\equiv g^2(k_{np})/(4\pi w_1)\]
reaches the value
\be\label{3.20}
\alpha_s(k\to0)=3.35\ee

\section{Gluon condensate}
\setcounter{equation}{0}

Let us now turn to the gluon condensate as
described by the expectation value of a suitably smeared composite
operator $\sim <F_{\mu\nu}F^{\mu\nu}>$. This expectation value
may not vanish despite a vanishing
ground state value for the gauge field
$<A^\mu>=0$. The formalism for the introduction of composite
fields in the flow equations is described in \cite{Ell},
and we use it here in the version without
a separate infrared cutoff for the
composite field $\chi$. Essentially, this amounts to the
introduction of an identity of the type
\be\label{6.1}
1=const\int {\cal D}\chi\exp\left[-\int d^dx\frac{1}{2ak^4_\chi}\left\{
k^3_\chi\chi-\frac{1}{4}F^z_{\mu\nu}F^{\mu\nu}_z\right\}^2\right]\ee
into the functional integral defining the effective average action
$\Gamma_{k\chi}$. The scale $k_\chi$ and the parameter $a$
should be chosen conveniently such that a maximum of the effects of
higher invariants can be described by the dynamics of the new scalar
singlet $\chi$. Typically $k_\chi$ should be a scale where the
influence of $w_2$ and $w_3$ on the running of $w_1$ (or $g^2$)
becomes important. The expectation value $<\chi>$ for $k\to0$ is directly
related  \cite{Ell} to the expectation value of the composite
operator $F_{\mu\nu}F^{\mu\nu}$
\be\label{4.1a}
<\chi>=\frac{1}{4}k_\chi^{-3}<F^z_{\mu\nu}F^{\mu\nu}_z>.\ee
The operator appearing on the r.h.s. is
regulated by smearing over distances $\sim k_\chi^{-1}$. For scales
$k<k_\chi$ we now have to deal with the coupled system of $A_\mu$ and
$\chi$. The exact flow equation has the same general
structure as before, except for the
extension of $\Gamma_k^{(2)}$ to the second functional derivative
including also the new field $\chi$. Nevertheless, since in the
formulation used here $R_k$ acts only on gauge fields and ghosts,
only the projection of $(\Gamma_k^{(2)}+R_k)^{-1}$ on this
restricted space enters the flow equations. Formally, the
components $R_{\chi\chi}$ and $R_{\chi A}$ vanish in
the generalized trace on the r.h.s. of the flow equation.

Including the composite field, we extend our truncation for the average
action (\ref{2.1}) by the addition of a term
\be\label{6.2}
\Gamma_k[A,\chi]=\int d^dx\{V_k(\frac{1}{4}F^z_{\mu\nu}F_z^{\mu\nu},
\chi)+\frac{1}{2}Z_{\chi,k}(\chi)\partial^\mu\chi\partial_\mu\chi\}\ee
At the scale $k_\chi$, eq. (\ref{6.1}) implies
the ``initial values''  \cite{Ell}
\ba\label{6.3}
V_{k\chi}(\frac{1}{4}F^z_{\mu\nu}F^{\mu\nu}_z,\chi)&=&W_{k\chi}
(\frac{1}{4}
F^z_{\mu\nu}F_z^{\mu\nu})+\frac{k_\chi^2}{2a}\chi^2,
\nonumber\\
&-&\frac{1}{4ak_\chi}
\chi F^z_{\mu\nu}F^{\mu\nu}_z+
\frac{1}{32ak^4_\chi}\left(F^z_{\mu\nu}F^{\mu\nu}_z
\right)^2\nonumber\\
&&\nonumber\\
Z_{\chi,k_\chi}(\chi)&=&0\ea
The expectation value $<\chi>$ for $<A_\mu>=0$
 corresponds to the minimum
of the effective scalar potential
\be\label{6.4}
U_k(\chi)=V_k(0,\chi)\ee
for $k\to0$. At the scale $k_\chi$ this potential is
simply quadratic, $U_{k_\chi}=\frac{1}{2a}k_\chi^2
\chi^2$, and has its minimum for $\chi=0$. We will be interested in the
change of shape of $U_k$ as $k$ flows towards zero.

As in the preceding
sections we will be concerned with configurations of space-independent
static magnetic fields where
$\theta=\frac{1}{4} F^z_{\mu\nu}F^{\mu\nu}_z$ is a
constant. We note that $V_{k_\chi}(\theta,\chi=0)$ differs from
$W_{k_\chi}(\theta)$ by the subtraction of terms $\sim\theta^2,
\theta^3$ and $\theta^4$. For all $k$, the original function $W_k(\theta)$
can be recovered by solving the field equation for $\chi, \frac{\partial
V}
{\partial\chi}(\theta,\chi_0(\theta))=0$, and inserting
$\chi_0(\theta)$ into $V$, i.e. $W(\theta)\equiv
V(\theta,\chi_0(\theta))$.
Nevertheless, the introduction of $\chi$ effectively extends the
truncation
and therefore results in modified flow equations for $W_k(\theta)$. In
particular, we can now expand $V_k(\theta,\chi)$ in powers of $\theta$ for
arbitrary constant $\chi$.

Let us next derive the evolution equation for the scale dependence of
$V_k$.
This is done by evaluating the general flow equation for constant $\chi$
and
$B$, inserting on the r.h.s. the truncation (\ref{6.2}).
We need the inverse propagator
$\Gamma^{(2)}_k$ for this ``background'' configuration. The components
$(\Gamma_k^{(2)})_{A_\mu A_\nu}$  remain the same as in the previous
sections,
if we simply replace $W_k(\theta)$ by $V_k(\theta,\chi)$. The pure scalar
 piece  is also easily obtained\footnote{$\tilde Z_\chi(\chi)$ is
related to $Z_\chi(\chi)$ and its derivative in a simple way.}
\be\label{6.5}
(\Gamma_k^{(2)})_{\chi\chi}=\frac{1}{2}\frac{\partial^2 V_k
(\theta,\chi)}{\partial\chi^2}+\tilde Z_{\chi,k}(\chi)(-\partial^2).\ee
We note, however, that the scalar fluctuations can influence the flow
equation only indirectly through the off diagonal  piece $(\Gamma^{(2)}_k)
_{\chi A_\mu}$. Using
\be\label{6.6}
\frac{\delta\Gamma_k}{\delta A^z_\mu(x)}=\int d^dz\frac{\delta\theta(z)}
{\delta A^z_\mu(x)}\frac{\partial V_k}{\partial\theta}(z)=
(D_\nu F^{\mu\nu})_z(x)\frac{\partial V_k}{\partial \theta}(x)\ee
we find  for the constant magnetic field background
\be\label{6.7}
\frac{\delta^2\Gamma_k}{\delta A^z_\mu(x)\delta\chi(y)}=
(D_\nu F^{\mu\nu})_z(x)\frac{\delta}{\delta\chi(y)}\frac{\partial V_k}{
\partial\theta}(x)=0.\ee
For this configuration we therefore find a particularly simple flow
equation:
Since $(\Gamma^{(2)}_k+R_k)$ is block diagonal and  $R_{\chi\chi}=0$,
it reduces to the same flow equation as before, with $W_k(\theta)$
replaced
by $V_k(\theta,\chi)$!

Introducing according to (\ref{3.17})
\be\label{6.8}
v_k(\vartheta,\chi)=g^2 k^{-d} V_k(\theta,\chi)\ee
the evolution equation for $v_k$ is given by (\ref{3.18}),
with $w_k(\vartheta)$ replaced by $v_k(\vartheta,\chi)$ and dots denoting
partial derivatives with respect to $\vartheta$ at fixed $\chi$.
Typically we will  choose $k_\chi=k_{np}$  such that $g^2(k)\equiv g^2
(k_{np})$. Expanding as before in  cubic order in $\vartheta$
\be\label{6.9}
v_k(\vartheta,\chi)=u(\chi)+v_1(\chi)\vartheta+\frac{1}{2} v_2
(\chi)\vartheta^2+\frac{1}{6} v_3(\chi)\vartheta^3\ee
the flow equations for $v_1(\chi),v_2(\chi)$ and $v_3(\chi)$ are given by
(\ref{4.9}) and (\ref{5.8}), with the replacement $w_i\to v_i(\chi)$.
The evolution of $u(\chi)$ follows from (\ref{3.18}) for
$\vartheta=0$ and $d=4$, i.e.
\be\label{6.10}
\frac{\partial}{\partial t} u =-4 u+\frac{3\zeta(3)}{8\pi^2}
(N^2-1)\frac{g^2(k_{np})}{v_1^2}.\ee
The system of equations (\ref{6.10}), (\ref{4.9}), and
(\ref{5.8}) still remains a complicated system of four coupled partial
differential equations for the functions $u(\chi,t),\  v_1(\chi,t)$,\\
$v_2(\chi,t)$ and  $v_3(\chi,t)$ which  depend on two variables.

In order to gain some intuition about  this system let us first  look at
the differential equation for the scalar potential  $U_k(\chi)$
and its derivatives
\ba
\frac{\partial}{\partial t} U&=&\frac{3\zeta(3)}{8\pi^2}(N^2-1)k^4
v_1^{-2}\label{6.11}\\
\frac{\partial}{\partial t}\left(\frac{\partial U}{\partial\chi}
\right)&=&-\frac{3\zeta(3)}{4\pi^2}(N^2-1) k^4 v_1^{-3}
\frac{\partial v_1}{\partial\chi}\label{6.12}\\
\frac{\partial}{\partial t}\left(\frac{\partial^2 U}{\partial \chi^2}
\right)&=&\frac{3\zeta(3)}{4\pi^2}(N^2-1) k^4 v_1^{-4}\left[3
\left(\frac{\partial v_1}{\partial\chi}\right)^2-v_1
\frac{\partial^2 v_1}{\partial \chi^2}\right]\label{6.13}.\ea
At the scale $k_\chi$ the initial values for $v_1$ and $\partial
v_1/\partial\chi$ are given by
\ba\label{6.14}
v_{1_{|k_\chi}}&=&w_{1|k_\chi}-\frac{g^2(k_{np})}{\bar g^2 ak_\chi}\chi
\nonumber\\
\frac{\partial v_1}{\partial\chi}_{|k_\chi}&=&
-\frac{g^2(k_{np})}{\bar g^2ak_\chi}\ea
whereas $\partial^2 v_1/\partial \chi^2_{|k_\chi}$
vanishes. The r.h.s. of
eq. (\ref{6.12}) starts therefore positive and
induces a negative linear term
in $U$ at $\chi=0$, resulting in a minimum at $\chi_0(k)>0$ for
$k<k_\chi$. Simultaneously, the positive
r.h.s. of eq. (\ref{6.13}) leads to
a decrease of the mass term for $\chi$. One
can follow the $k$-dependence
of the minimum value $\chi_0(k)\
(\chi_0\equiv\chi_0(\vartheta=0))$ using the
identity
\be\label{6.15}
\frac{\partial \chi_0}{\partial t}=-
m_\chi^{-2}\frac{\partial}{\partial t}
\left(\frac{\partial U_k}{\partial \chi}\right)(\chi_0)\ee
where we have defined the scalar mass term
\be\label{6.16}
m_\chi^2\equiv\frac{\partial^2 U}{\partial\chi^2}(\chi_0).\ee
Inferring the r.h.s. from eqs. (\ref{6.12}), (\ref{6.13})
evaluated at $\chi=\chi_0$ we conclude that the running of $\chi_0$ stops
for small $k$ due to the  factor $k^4$, provided $m^2_\chi$ and $v_1$
remain nonzero for $k\to 0$.

Let us next consider an expansion in powers of $\vartheta$ for $\chi$
at the minimum $\chi_0(k)$ and compare
\ba\label{6.17}
v_n&\equiv& v_n(\chi_0)=\frac{\partial^n}{\partial \vartheta^n}v(
\vartheta,\chi_0(\vartheta=0))_{|\vartheta=0}\nonumber\\
w_n&\equiv &\frac{\partial^n}{\partial\vartheta^n} w(\vartheta=0)
=\frac{d^n}{d\vartheta^n} v(\vartheta,\chi_0(\vartheta))
_{|\vartheta=0}\ea
Whereas $w_1=v_1$ holds for all $k$ the corresponding
relation for $w_2$ reads
\be\label{6.17b}
w_2=v_2-\frac{\partial^2 u}{\partial \chi^2}(\chi_0)\frac{\partial\chi_0(
\vartheta)}{\partial\vartheta}_{|\vartheta=0}.\ee
For positive $\chi_0$ and negative $\frac
{\partial v_1(\chi)}{\partial\chi}$
we observe $v_1\equiv v_1(\chi_0)<v_1(0)$. Due to the running $\chi_0$ the
flow equation for $v_1$ obtains an additional contribution
\be\label{6.18}
\frac{\partial}{\partial t} v_1=\frac{g^2(k_{np})}{8\pi^2}
\left(\frac{11 N}{3}-H_4\frac{v_2}{v_1^3}\right)+
\frac{\partial v_1(\chi)}{\partial\chi}_{|\chi_0}
\frac{\partial}{\partial t}
\chi_0,\ee
and similar for all $v_n\equiv v_n
(\chi_0)$ where a term $\frac{\partial
v_n(\chi)}{\partial\chi}_{|\chi_0}\frac{\partial\chi_0}{\partial t}$
should be added on the r.h.s. of the flow equations.

It is instructive to compare the flow equation for $v_1$
to the one without a condensate (\ref{4.9}).
In the vicinity of $k_\chi$ we can use the initial
values (\ref{6.16}), together with
\ba\label{6.19}
&&v_{2_{|k_\chi}}=w_{2_{|k_\chi}}+\frac{g^2(k_{np})}{a\bar g^4}
\nonumber\\
&&v_{3_{|k_\chi}}=w_{3_{|k_\chi}}
\nonumber\\
&&m^2_{\chi|k_\chi}=\frac{k_\chi^2}{a}\ea
As compared to eq. (\ref{4.9}) the difference between
$v_2$ and $w_2$ subtracts from the r.h.s. of (\ref{6.18}) a piece
$g^4(k_{np})H_4/(8\pi^2\bar g^4av_1^3)$. On the other hand, the
contribution $\sim \frac{\partial v_1}{\partial\chi}
\partial\chi_0/\partial t=g^4(k_{np})(H_4-r^4_1)/(8\pi^2\bar g
^4av^3_1)$ almost cancels this piece except for the negative
contribution $\sim r^4_1$. The flow of $w_1$ is therefore somewhat
slower than the one corresponding to (4.9).

We also need flow equations for the quantities
\be\label{6.20}
y_n=\frac{\partial v_n(\chi)}{\partial
\chi}(\chi_0)=\frac{\partial}{\partial\chi}\frac{\partial^n}{\partial
\vartheta^n}v_{|\vartheta=0,\chi=\chi_0}\ee
which appear on the r.h.s. of the flow equations for $v_n$, $\chi_0$
and $m_\chi^2$. They obtain by partial differentiation of (4.9)
and (5.8) with respect to $\chi$:
\be\label{6.21}
\frac{\partial}{\partial t}y_1=\frac{H_4g^2(k_{np})}{8\pi^2}\left(
3\frac{v_2y_1}{v_1^4}-\frac{y_2}{v^3_1}\right)\ee
\ba\label{6.22}
\frac{\partial}{\partial t}y_2&=&4y_2+\frac{g^2(k_{np})}{4\pi^2}
\Bigl\{\frac{29}{10}\tau_2r^{4,2}_0v_1y_1
+(5r_1^4+\frac{9}{2}(N^2-1)r^4_2)(2\frac{v_2y_2}{v_1^4}-4\frac
{v_2^2y_1}{v_1^5})\nonumber\\
&&-(r^4_1+\frac{3}{2}(N^2-1)r^4_2)(\frac{y_3}{v_1^3}-3\frac{v_3y_1}
{v_1^4})\Bigr\}\ea
\ba\label{6.23}
\frac{\partial}{\partial t}y_3&=&8y_3+\frac{g^2(k_{np})}{16\pi^2}
\Bigl\{\frac{548}{105}\tau_3r^{4,3}_0v_1^3y_1
+\frac{174}{5}\tau_2r^{4,2}_0(v_1y_2+v_2y_1)\nonumber\\
&&+6r^4_1(17\frac{v_2y_3}{v_1^4}+17\frac{v_3y_2}{v_1^4}-68\frac
{v_2v_3y_1}{v_1^5}
-105\frac{v_2^2y_2}{v_1^5}+175\frac{v_2^3y_1}{v_1^6}-\frac{y_4}{v_1^3}
+3\frac{v_4y_1}{v_1^4})\nonumber\\
&&+6r^4_2(N^2-1)(9\frac{v_2y_3}{v_1^4}+9\frac{v_3y_2}{v_1^4}-36\frac
{v_2v_3y_1}{v_1^5}\nonumber\\
&&-36\frac{v_2^2y_2}{v_1^5}+60\frac{v_2^3y_1}{v_1^6}-\frac{y_4}{v_1^3}
+3\frac{v_4y_1}{v_1^4})\Bigr\}\ea
As before, we truncate by neglecting terms $\sim v_4$ and
$\sim y_4$ and we further omit $\partial^2v_1/\partial\chi^2(\chi_0)$
and a term $\sim\frac{\partial^3U}{\partial\chi^3}(\chi_0)\frac{\partial
\chi_0}{\partial t}$  in the flow equation for $m_\chi^2$ (\ref{6.13}):
\be\label{6.24}
\frac{\partial}{\partial
t}m_\chi^2=\frac{9\zeta(3)}{4\pi^2}(N^2-1)k^4v_1^{-4}y^2_1\ee
We then end up with a closed system of differential equations for
the eight
functions $v_1,v_2,v_3, \chi_0,m_\chi^2,y_1,y_2,y_3$. This system can
be solved numerically. The initial conditions for $y_i$ at the scale
$k_\chi$ follow from (\ref{6.3}):
\be\label{6.25}
y_{1_{|k_\chi}}=-\frac{g^2(k_{np})}{\bar g^2ak_\chi},\quad
y_{2_{|k_\chi}}=0,\quad y_{3_{|k_\chi}}=0\ee
We note that the choice of $\bar g^2$ is
arbitrary (it concerns only the overall normalization of $\theta$) and
it seems convenient to take $\bar g^2=g^2(k_{np})$.

In order to make some quantitative comparison we compute in addition the
trace of the energy momentum tensor $T^\mu_\mu$. In QCD its nonzero
value is related to the running of the coupling constant and it is
usually
quoted as $T^\mu_\mu=-0.014$ GeV$^4$.
In our context the part of $T^\mu_\mu$ induced by the gluon
condensate can be expressed in terms of the
effective potential for $\chi$,
\be\label{6.26}
T^\mu_\mu=4[U(\chi_0)-U(0)]\ee
Its value at $k=0$ can be computed from the flow equation (\ref{6.11})
\be\label{6.27}
\frac{\partial}{\partial t}T^\mu_\mu=4\frac{\partial}{\partial
t}U_k(\chi_0)=\frac{3\zeta(3)}{2\pi^2}(N^2-1)
k^4[v_1^{-2}-v_1(\chi=0)^{-2}]\ee
with $T_\mu^\mu(k_\chi)=0$. We use a linear approximation
for $v_1(\chi)$, i.e. $v_1(\chi=0)=v_1-y_1\chi_0$. As
an alternative, we can estimate $T^\mu_\mu$ in the
approximation of a quadratic potential $U=\frac{1}{2}m_\chi^2
(\chi-\chi_0)^2$, i.e.
\be\label{6.28}
T^\mu_\mu=-2m^2_\chi\chi^2_0\ee

We have solved the system of differential equations for $v_1, v_2, v_3,
y_1, y_2, y_3, \chi_0$ and $m_\chi^2$  numerically.
More precisely, we have first solved for $k>k_{np}$ the system of
flow equations of sect. 3 without composite fields. Here we have
started at $k=4$ GeV with a gauge
coupling adjusted in such a waythat the scale $k_{np}$
coincides with a similarly defined scale in terms
of the coupling in the $\overline{MS}$-scheme
with three light quark flavours:
\[\beta_{g^2}/g^2_{\overline{MS}}
(k_{np})\simeq-1.5.\]
 In practice we set
\[k_{np}=2.11\ \Lambda^{\overline
{MS}}_{\rm QCD}=600\ {\rm MeV}\]
(for $\Lambda^{\overline
{MS}}_{\rm QCD}=285$ MeV the two-loop confinement
scale). At the scale $k_{np}$ we also have
introduced the composite field
according to eq. (\ref{6.3}), with $k_\chi=k_{np}$. The results depend
somewhat on the parameter $a$ that enters the initial
values (\ref{6.14}), (\ref{6.19}), (\ref{6.25}).
We consider explicitly two values of $a$,
\be\label{6.29}
a=-\frac{A}{\bar g^2w_2(k_\chi)},\qquad A=2(1.1)\ee
where the value in brackets corresponds to a situation for which the system
becomes numerically unstable. For $A=3$ we obtain similar values
as for $A=2$. We consider the values for $A=2$ as our
best estimate and take the values for $A=1.1$ as an indication
of the error.

One obtains for $k\to0$ the same fixed point values for
$v_1, v_2, v_3$ and $\alpha_s$ as found before without composite
operators (end of the preceding section). This does not change for
larger values of $a$. For $k=0$ the expectation value
$\chi_0$ reaches the value
\[\chi_0=0.65\ (1.45)\  {\rm GeV}\]
for which
\[<\chi>/\Lambda_{\rm QCD}=2.3\ (5.1)\]
and the ``$F^2$'' condensate equals
\[<F^z_{\mu\nu}F^{\mu\nu}_z>\equiv4k^3_\chi<\chi>=[3.1\ (3.7)
\Lambda_{\rm QCD}]^4\]
Here the first and the second figure always refers
to $A=2$ and $A=1.1$, respectively. From
the expressions (\ref{6.26}), (\ref{6.28}) for $T_\mu^\mu$ one finds
\[T^\mu_\mu=-0.015\ (-0.039)\ {\rm GeV}^4\ {\rm and}\ T^\mu_\mu=-0.0145\ (-0.0295)\
{\rm GeV}^4,\]
respectively.
Comparing with the value $T^\mu_\mu=-0.014\ \GeV^4$ from
QCD sum rules the agreement is almost perfect. Since we have made
no attempt to define the renormalization scheme precisely and
quarks are neglected, there is, however, a considerable scale
ambiguity on top of the truncation errors.
The mass term turns out as $m_\chi
=130\ (83)$ MeV. This should, however, not yet be associated with
the mass of a scalar glueball\footnote{The quantity
$m_\chi^{-1}$ should also not be confused with the correlation
length for field strength fluctuations \cite{Dosch}.}.
The latter needs knowledge about the
kinetic term of the corresponding operator and could be defined
as $m_{gb}=m_\chi Z_\chi^{-1/2}$. It seems possible to compute $Z_\chi$
from the solution of its flow equation, starting at $k_\chi$ with
$Z_\chi=0$.

We do not think that the precise
numbers should be taken too seriously.
Even within our truncation one could  solve the
partial differential equation for $V_k(\theta,
\chi)$ without a polynomial truncation. One may also
investigate the dependence of the results on the
parameters $k_\chi$ and $a$. This would also give a rough
estimate of the truncation uncertainties since without
truncations the final values of the condensates should be
independent of $k_\chi$ and unique for a given definition of the
composite operator. Nevertheless, the fact that expectation
values come out in a reasonable order of magnitude seems
to indicate that our approach offers a possible
perspective for a quantitative
understanding of the gluon condensate.

\section{Conclusions and outlook}
\setcounter{equation}{0}
In this paper we have approximated the exact nonperturbative evolution
equation for Yang-Mills theories \cite{reu} by a truncation where
the effective action $\Gamma_k$
is given as a function $W_k$ of
$\theta=\frac{1}{4}F_{\mu\nu}F^{\mu\nu}$.
More precisely, we have investigated the
effective action for constant magnetic fields $B$
in dependence on an infrared cutoff $k$ and solved
the corresponding flow equation for $k\to0$. Our
ansatz is general enough
to allow for a  ground state value of $B$ different from zero. This
would correspond to an absolute
minimum of $W_k(\theta)$ for $\theta_0>0$ in the limit where the
infrared cutoff $k$ vanishes.
Polynomial approximations of $W_k$ in order
$\theta$ and $\theta^2$ seem to indicate that the
minimum of $W_{k\to0}$
occurs for $\theta_0>0$, but are obviously
insufficient since they lead to a function $W(\theta)$
which is not bounded from below. The minimal realistic truncation in
order $\theta^3$ leads to the conclusion that the minimum of $W_k(\theta)$
stabilizes at $\theta_0=0$ for all $k$. From an inspection
of the general structure of the flow equation for $W_k(\theta)$
it seems likely that this feature is not an artifact of the
polynomial approximation but rather a property of the partial
differential equation for $W_k(\theta)$.

For a description of gluon condensation we have,
in addition, introduced a gauge singlet field
$\chi$ which is associated to a suitably regularized composite
operator $F_{\mu\nu}F^{\mu\nu}$.
Our computation of the effective potential for $\chi$ indicates
a nonvanishing expectation value of $\chi$, related to
a nonvanishing expectation value $<F_{\mu\nu}F^{\mu\nu}>
\not=0.$ A first numerical investigation gives a value for
the condensate $<F_{\mu\nu}F^{\mu\nu}>\approx[3.1
\Lambda_{\rm QCD}]^4$ where $\Lambda_{\rm QCD}$ is the two-loop
confinement scale
in the $\overline{MS}$ scheme with three light flavors.
The precise meaning of the operator
$F_{\mu\nu}F^{\mu\nu}$ needs to be worked out - in the present
version it roughly corresponds to $F_{\mu\nu}\hat\theta
(k^2_\chi+D^2[A])F^{\mu\nu}$
with $D^2$ the covariant Laplacian and $k_\chi\approx 2.1\ \Lambda_{
\rm QCD}$. Here $\hat\theta$ stands for a rather sharp
cutoff (similar to the usual $\theta$-function) such that
high eigenvalues of $-D^2$ do not contribute to the
regularized operator. Even though we do not claim quantitative
precision of our estimate we find it remarkable that the
solution of relatively simple flow equations leads to a
reasonable order of magnitude of the condensate!

For higher precision, and, in particular, for a possible comparison
with more phenomenological approaches such as QCD sum rules
or the stochastic QCD vacuum several shortcomings of the
present truncation should be overcome. We list a few which seem
to us particularly important:

(i) In the present truncation the momentum dependence of
the inverse gluon propagator is always approximated by $Z_kq^2$,
with $Z_k=W_k'(0)$. For $k$ in the vicinity of $\Lambda_{\rm QCD}$
this is certainly not a very accurate approximation, even though
the flow equations involve effectively only a small momentum
range $q^2\approx k^2$ for a given scale $k$.
A more reasonable approximation for the term
quadratic in the gauge field $A_\mu$ would be of the sort
\be\label{7.1}
\frac{1}{4}\int d^4xF^z_{\mu\nu}Z_k(-D^2[A])F^{\mu\nu}_z\ee
with $Z_k$ depending on the covariant Laplacian in the adjoint
representation\footnote{The exact propagator will, in addition,
also receive corrections from the modified gauge-fixing term
$\hat\Gamma_{gauge}$ discussed in Appendix A.}.
For the functional form of $Z_k(q^2)$ we expect
for large $q^2$ a $k$-independent positive function $Z(q^2)$. In
lowest order the logarithmic dependence of
$Z$ on $q^2$ should be determined
by the one-loop $\beta$-function for the running gauge coupling. Indeed,
external momenta of the gluons act as an independent infrared cutoff. For
$k^2\ll q^2$ the running of $Z_k(q^2)$ with $k$ should essentially
stop whereas it is given by the one-loop $\beta$-function
for $k^2\grgl q^2$.
In contrast, our truncation identifies $Z_k(q^2)$ with a
constant\footnote{This would lead
to unphysical problems for vanishing or
negative $W_k'$. With (\ref{7.1}) the quantity $Z_k(0)$ could turn
negative without affecting the high momentum behavior of $Z_k(q^2)$.
With such a truncation the high momentum modes
are always stable if the momentum dependence of the propagator is
properly taken into account. This implies positive $u$ for $q^2\to\infty$
and no ultraviolet problem can appear for the exponential cutoff
(\ref{1.2}).}
$Z_k(q^2=0)\equiv W_k'(\theta=0)$. In the truncations of sects. 3 and 4,  
$W_k'(\theta)$
remains positive for all $\theta$ such that the use of
$Z_k=W_k'(0)$ induces quantitative inaccuracy, but no
qualitative problems.

ii) We observe that the spectrum of small fluctuations around the
constant magnetic field configuration (\ref{2.7}) lacks Euclidean
$SO(4)$ rotation symmetry. This is not surprising since $F_{\mu\nu}$
singles out two of the space directions. In addition, the spectrum is
partially discrete - continuity exists only with respect to the
transversal momentum. The lack of full rotation symmetry and the partial
discreteness of the spectrum are actually not unrelated:
For an $SO(4)$
symmetric spectrum the continuity in two momentum directions must
extend to all momentum directions. One expects then a spectrum
with a few separate particles, each of them having $\Gamma_k^{(2)}$
depending continuously on a generalized $SO(4)$ invariant of the type
$q_\mu q^\mu$. The lack of full rotation symmetry for the fluctuation
spectrum in a given background is not a problem. Nevertheless, it would
be interesting to check the validity of approximations by
investigating background configurations with a different symmetry.
In ref. \cite{reuwe}
we have proposed a candidate for a configuration with
generalized $SO(4)$ symmetry for $d=4$
and gauge group $SU(N),\ N\geq 4$. Unfortunately,
no realistic candidate for the gauge group $SU(3)$ has been found
up to now. It would be very interesting to perform an analysis similar
to the one presented here for $SU(4)$, using the $SO(4)$ symmetric
configuration instead of (\ref{2.7}).

For the gauge group $SU(3)$ an interesting alternative
configuration with constant colour-magnetic field is given by
\be\label{7.2}
A_i^z=\left\{ \begin{array}{clc}
a \delta^z_i& {\rm for}&
i=1,2,3 \\
0& {\rm for}& i>3\end{array}\right.\ee
(with $i=1,2,3$ corresponding to the spacelike indices in
Minkowski space). It is invariant under the diagonal subgroup
$SO(3)$ of three-dimensional space rotations and global gauge
transformations (with generators $T_1, T_2, T_3$ forming an $SU(2)$
subgroup of $SU(3)$). The field strength corresponds to a constant
colour-magnetic field
\ba\label{7.3}
&&F_{zij}=\left\{ \begin{array}{cl}
\bar g a^2\epsilon_{zij}& {\rm for} \ i,j=1,2,3 \\
0 &{\rm otherwise}\end{array}\right.\nonumber\\
&&F^{\mu\nu}_zF^z_{\mu\nu}=2B^2=6\bar g^2a^4\ea
Nevertheless, $F_{\mu\nu}$ is not covariantly constant
$(i,j,k=1,2,3)$
\be\label{7.4}
(F^{ij}_{\ \ ;k})_z=-\bar g^2a^3(\delta^i_k\delta^j_z-\delta
^j_k\delta^i_z)\ee
\be\label{7.5}
(-D^2[A])^{yz}F^z_{ij}=2\bar g^2a^2F^y_{ij}\ee
and this configuration allows to explore a nontrivial
momentum dependence of the gluon propagator as in (\ref{7.1}). For
the characteristic properties of the spectrum of fluctuations
around the configuration (\ref{7.2}) it is sufficient to restrict
the discussion to $d=3$. With the truncation (\ref{2.1}) we find
that the spectrum is continuous.
For large momenta it turns out positive semi-definite for all $B$,
even for $k=0$. This contrasts with the spectrum for the background
(\ref{2.7})! For a three-dimensional theory (high temperature
field theory) the configuration (\ref{7.2}) is an interesting
ground state candidate. For the four-dimensional theory a study
of $\Gamma_k$ as a function of $a$ should reveal additional
aspects of the functional form of $\Gamma_k$.

In general, an investigation of the functional dependence of $\Gamma_k$
on various distinct configurations would greatly enhance the
robustness of our results. Qualitative and quantitative results
should be independent of the configuration chosen. The
use of different configurations should therefore permit an
estimate of the truncation uncertainties. In addition, different
configurations project on different invariants and the use of
several configurations would give much more detailed information
about the full functional form of the effective action $\Gamma_k$.

iii) The present treatment of composite operators is still rather
rough. One would prefer to work directly with a smooth definition
of a regularized operator, as, for example,
\be\label{7.6}
\chi\hat{=}\frac{1}{4k^3_\chi}F_{\mu\nu}\exp(D^2[A]/k^2_\chi)F^{\mu\nu}\ee
This would permit a more direct comparison of the expectation value
of $\chi$ with a corresponding weighted momentum integral
of the two-point function for $F_{\mu\nu}$ and therefore with
lattice results or phenomenological estimates. The general formalism
for the treatment of this type of operators is known \cite{Ell}, but in
practice it will require a more detailed study of the momentum
dependence of $\Gamma_k$, beyond the truncation $\sim W_k(\theta)$. In
fact, for the configuration with constant $B$ used in this paper
$D^2[A]F^{\mu\nu}$ vanishes. At the present stage we can therefore
not distinguish between different versions of a regularized
$F_{\mu\nu}F^{\mu\nu}$ operator. (This is the
reason why we cannot specify the function $\hat\theta$ mentioned
above.) In order to resolve this issue, one needs an
investigation of different configurations as described in ii).

In summary, our first attempt to investigate the phenomenon of
gluon condensation with the help of nonperturbative flow equations
is encouraging. The simple configuration (\ref{2.7}) and the
truncation (\ref{2.1}) give a qualitatively interesting picture: Whereas
the ground-state value of the gauge field $A_\mu$ vanishes,
a composite operator $\sim F_{\mu\nu}F^{\mu\nu}$ develops a
non-vanishing vacuum expectation value. For the moment our analysis
is too rough for a detailed comparison with quantities that can
be determined in lattice QCD simulations or by phenomenological
estimates as QCD sum rules. We hope that a future extension
of our investigation will allow for such a comparison. We find it
remarkable that already a relatively simple scheme
which starts from first principles and leads to a
quantitative estimate of the gluon condensate.

\bigskip
\noindent{\bf Acknowledgement:} We would like to thank D. Jungnickel
for a first numerical solution of the differential
equations of sect. 3.

\section*{Appendix A}
\renewcommand{\theequation}{A.\arabic{equation}}
\setcounter{equation}{0}

In this appendix we derive the exact
form of the evolution equation for $\Gamma_k$ as well as the modified Ward-Takahashi
or Slavnov-Taylor identities and ``background field
identities'' which it satisfies.  We start
from the scale dependent generating functional
\ba\label{AA.1}
\exp W_k[K^z_\mu,\sigma^z,\bar\sigma^z;\bar\beta^z_\mu,\bar\gamma^z;\bar
A^z_\mu]&=&\int {\cal D} {\cal A} {\cal D} C{\cal D}\bar C\exp-\left\{
S[{\cal A}]+\Delta_kS\right.\nonumber\\
\left. +S_{\rm gf}+S_{\rm ghost}+S_{\rm source}\right\}&=&\int {\cal
D}\phi\exp-S_{tot}\ea
where $S[{\cal A}]$ denotes the gauge invariant classical action, and
\ba\label{AA.2}
\Delta_kS&=&\frac{1}{2}\int d^dx({\cal A}-\bar A)^y_\mu R_k(\bar
A)^{yz}_{\mu\nu}({\cal A}-\bar A)^z_\nu\nonumber\\
&&+\int d^d x\bar C^y R_k(\bar A)^{yz} C^z\ea
is the infrared cutoff for the gauge field fluctuation $a\equiv {\cal
A}-\bar A$ and for the Faddeev-Popov ghosts $C$ and $\bar C$. Here
$R_k(\bar A)$ is a suitable cutoff operator which depends on $\bar A$
only. It may be chosen differently for the gauge field and for the ghosts.
Furthermore
\be\label{AA.3}
S_{\rm gf}=\frac{1}{2\alpha}\int d^d x
\left[ D_\mu(\bar A)^{yz}({\cal A}-\bar A)^z_\mu\right]^2\ee
is the background gauge fixing term and
\be\label{AA.4}
S_{\rm ghost}=-\int d^dx \bar C^y\left(D_\mu(\bar A) D_\mu({\cal
A})\right)^{yz} C^z\ee
is the corresponding ghost action \cite{abb}.
The fields ${\cal A}-\bar A$, $\bar C$ and $C$ are
coupled to the sources $K,\ \sigma$ and $\bar\sigma$, respectively:
\ba\label{AA.5}
S_{\rm source}&=&-\int d^d x\Bigl\{ K^z_\mu({\cal A}^z_\mu-\bar A_\mu^z)+
\bar\sigma^z C^z+\sigma^z\bar C^z\nonumber\\
&&+\frac{1}{g}\bar\beta^y_\mu D_\mu({\cal A})^{yz} C^z+\frac{1}{2}
\bar\gamma^w f^{wyz} C^y C^z\Bigr\}.\ea
(In this appendix we use throughout $g$ instead of $\bar g$.)
We also included the sources $\bar\beta$ and $\bar\gamma$ which couple to
the BRS-variations of ${\cal A}$ and of $C$, respectively. In fact,
$S+S_{gf}+S_{\rm ghost}$ is invariant under the BRS transformation
\ba\label{AA.6}
\delta {\cal A}^w_\mu&=&\frac{1}{g}\varepsilon D_\mu({\cal A})^{wz}
C^z\nonumber\\
\delta C^w&=&-\frac{1}{2}\varepsilon f^{wyz}C^y C^z\nonumber\\
\delta \bar C^w&=&\frac{\varepsilon}{\alpha g} D_\mu (\bar A)^{wz}
({\cal A}_\mu^z-\bar A_\mu^z).\ea

Let us introduce the classical fields
\be\label{AA.7}
\bar a^z_\mu=\frac{\delta W_k}{\delta K_\mu^z},\
\xi^z=\frac{\delta W_k}{\delta \bar\sigma^z},
\ \bar \xi^z=\frac{\delta W_k}{\delta\sigma^z}\ee
and let us formally solve the relations
$\bar a=\bar a(K,\sigma,\bar\sigma;\bar\beta,\bar\gamma;\bar A),
\ \xi=\xi(...)$, etc., for the sources $K,\sigma$ and
$\bar\sigma:\ K=K(\bar a,\xi,\bar \xi;\bar\beta,\bar\gamma;\bar A),\
\sigma=\sigma(...),\ ...$. We introduce the
new functional $\tilde\Gamma_k$ as the Legendre transform of $W_k$ with
respect to $K,\sigma$ and $\bar\sigma$:
\ba\label{AA.8}
\tilde\Gamma_k[\bar a, \xi,\bar \xi;\bar\beta,\bar\gamma;\bar A]&=&
\int d^dx\{K^z_\mu\bar a^z_\mu+\bar\sigma^z \xi^z+
\sigma^z\bar\xi^z\}\nonumber\\
&&-W_k[K,\sigma,\bar\sigma;\bar\beta,\bar\gamma;\bar A].\ea
Apart from the usual relations
\be\label{AA.9}
\frac{\delta\tilde\Gamma_k}{\delta\bar a^z_\mu}=K^z_\mu,\qquad
\frac{\delta\tilde\Gamma_k}{\delta \xi^z}=-\bar\sigma^z,\qquad
\frac{\delta\tilde\Gamma_k}{\delta\bar \xi^z}=-\sigma^z\ee
we have also
\be\label{AA.10}
\frac{\delta\tilde\Gamma_k}{\delta\bar\beta^z_\mu}=-
\frac{\delta W_k}{\delta\bar\beta^z_\mu},\qquad
\frac{\delta\tilde\Gamma_k}{\delta\bar\gamma^z}=-
\frac{\delta W_k}{\delta\bar\gamma^z}\ee
where $\delta\tilde \Gamma/\delta\bar\beta$ is
taken for fixed $\bar a,\xi,\bar \xi$ and
$\delta W/\delta\bar\beta$ for fixed
$K,\sigma,\bar\sigma$ etc.

The effective average action $\Gamma_k$ is
obtained by subtracting the IR cutoff $\Delta_k S$,
expressed in terms of the classical fields,
from the Legendre transform $\tilde\Gamma_k$:
\ba\label{AA.11}
\Gamma_k[\bar a,\xi,\bar \xi;\bar\beta,\bar\gamma;\bar A]&=&
\tilde\Gamma_k[\bar a, \xi,\bar \xi;\bar\beta,\bar\gamma;\bar A]
-\frac{1}{2}\int d^dx\bar a^y_\mu R_k(\bar A)^{yz}_{\mu\nu}\bar
a^z_\nu\nonumber\\
&&-\int d^dx\bar \xi^y R_k(\bar A)^{yz}\xi^z.\ea
Frequently we shall use the field $A\equiv\bar A+\bar a$ (the classical
counterpart of ${\cal A}\equiv \bar A+a)$ and write correspondingly
\be\label{AA.12}
\Gamma_k[A,\bar A,\xi,\bar \xi;\bar\beta,\bar\gamma]\equiv\Gamma_k[A-
\bar A,\xi,\bar \xi;\bar\beta,\bar\gamma;\bar A].\ee
For $\xi=\bar \xi=\bar\beta=\bar\gamma=0$
one recovers the effective average action
$\Gamma_k[A,\bar A]$ introduced in ref. \cite{reu}.
Using the methods described there one finds the following exact evolution
equation governing its scale-dependence, with $t=\ln k$
\ba\label{AA.13}
&&\frac{\partial}{\partial t}\Gamma_k[A,\bar A,\xi,\bar
\xi;\bar\beta,\bar\gamma]=\frac{1}{2}\Tr_{xcL}
\left[\left(\Gamma^{(2)}_k+R_k(\bar A)\right)
^{-1}_{AA}\frac{\partial}{\partial t}R_k(\bar A)_{AA}\right]
\nonumber\\
&&-\frac{1}{2}\Tr_{xc}\left[\left(\left(\Gamma^{(2)}_k+R_k(\bar
A))\right)^{-1}_{\bar \xi \xi}-\left(\Gamma_k^{(2)}+R_k
(\bar A)\right)^{-1}_{\xi\bar\xi}\right)
\frac{\partial}{\partial t} R_k(\bar A)_{\bar \xi \xi}\right].\ea
Here $\Gamma^{(2)}_k$ is the Hessian of $\Gamma_k$
with respect to $A,\xi$ and $\bar \xi$ at fixed $\bar A,
\bar\beta$ and $\bar\gamma$ and $R_{kAA},R_{k\bar \xi\xi}$
are the infrared cutoffs introduced in (\ref{AA.2}).
(We use for the ghosts the convention $\left(\Gamma^{(2)}_{\bar
\xi\xi}\right)^{yz}=\frac{\delta^2\Gamma}{\delta \xi^z\delta\bar \xi^y},
\left(\Gamma_{\xi\bar\xi}^{(2)}\right)^{yz}=\frac{\delta^2\Gamma}
{\delta\bar\xi^z\delta\xi^y}$.)

It is clear from its construction that $\Gamma_k$ is invariant under
simultaneous gauge transformations of $A_\mu$ and
$\bar A_\mu$ and homogeneous transformations of $\xi,\bar
\xi, \bar\beta_\mu$ and $\bar\gamma$, i.e., $\delta
\Gamma_k[A,\bar A,\xi,\bar \xi;\bar\beta,\bar\gamma]=0$ for
\ba\label{AA.14}
\delta A^y_\mu&=&-\frac{1}{g}D_\mu(A)^{yz} \omega^z\nonumber\\
\delta\bar A^y_\mu&=&-\frac{1}{g} D_\mu(\bar A)^{yz} \omega^z\nonumber\\
\delta V^y&=&f^{ywz} V^w \omega^z,\ V\equiv \xi,\bar \xi,
\bar\beta_\mu,\bar\gamma.\ea
\vspace{3mm}

Next we turn to the Ward identities. By applying the transformations
(\ref{AA.6}) to the integrand of (\ref{AA.1}) one obtains from the BRS
invariance of the measure ${\cal D}\phi$
\be\label{AA.15}
\int{\cal D}\phi\delta_{BRS}\exp-S_{\rm tot}=0\ee
or
\ba\label{AA.15a}
&&\int d^dx\left\{ K^z_\mu\frac{\delta W_k}{\delta\bar\beta^z_\mu}
+\bar\sigma^z
\frac{\delta W_k}{\delta \bar\gamma^z}-\frac{1}
{\alpha g}\sigma^y D_\mu(\bar A)^{yz}
\frac{\delta W_k}{\delta K^z_\mu}\right\}\nonumber\\
&=&\int d^dx\left\{\left[\frac{\delta
W_k}{\delta\bar\beta^z_\mu}+\frac{\delta}{\delta\bar\beta^z_\mu}
\right]\left( R_k\frac{\delta W_k}{\delta K}\right)^z_\mu+
\frac{1}{\alpha g}\left( D_\mu(\bar A)\left[\frac{\delta W_k}{\delta
K_\mu}+\frac{\delta}{\delta K_\mu}\right]\right)^z\right.
\left(R_k\frac{\delta W_k}{\delta\bar\sigma}\right)^z\nonumber\\
&&\left.+\left[\frac{\delta W_k}{\delta\sigma^z}+
\frac{\delta}{\delta\sigma^z}
\right]\left(R_k\frac{\delta W_k}{\delta\bar\gamma}\right)^z\right\}\ea
with $(R_k\delta W_k/\delta\bar\gamma)^z\equiv R_k
(\bar A)^{zy}\delta W_k/\delta \bar\gamma^y$, etc.
Equation (\ref{AA.15a}) can be converted
to the following relation for the effective average action (\ref{AA.12}):
\be\label{AA.16}
\int d^dx\left\{
\frac{\delta\Gamma'_k}{\delta
A^z_\mu}\frac{\delta\Gamma'_k}{\delta\bar\beta^z_\mu}
-\frac{\delta\Gamma'_k}{\delta \xi^z}\frac{\delta\Gamma'_k}
{\delta\bar\gamma^z}
\right\}=\Delta^{(BRS)}_k\ee
where the symmetry-breaking contribution $\Delta^{\rm(BRS)}_k$ is given by
\ba\label{AA.16a}
\Delta_k^{(BRS)}&=&\Tr_{xcL}\left[R_k(\bar A)_{A_\mu
A_\nu}(\Gamma^{(2)}_k+R_k)^{-1}_{A_\mu\varphi}
\frac{\delta^2\Gamma'_k}{\delta \varphi \delta
\bar\beta_\nu}\right]\nonumber\\
&&-\Tr_{xc}\left[ R_k(\bar A)_{\bar \xi
\xi}\left(\Gamma^{(2)}_k+R_k\right)^{-1}_{
\xi\varphi}\frac{\delta^2\Gamma'_k}{\delta\varphi\delta\bar\gamma}
\right]\nonumber\\
&&-\frac{1}{\alpha g}\Tr_{xc} \left[ D_\mu(\bar
A)\left(\Gamma^{(2)}_k+R_k\right)^{-1}_{A_\mu \bar\xi} R_k(\bar A)_{\bar
\xi \xi}\right]
\ea
and where
\be\label{AA.17}
\Gamma'_k\equiv \Gamma_k-\frac{1}{2\alpha}\int d^d x\left[ D_\mu (\bar A)
(A_\mu-\bar A_\mu)\right]^2.\ee
Here $\varphi\equiv (A_\mu,\xi,\bar \xi)$ is summed over on the r.h.s. of
(\ref{AA.16a}) and the traces on the r.h.s. of (A.18)
involve an $x$ integration as well as a sum over the
suppressed  index $z$. In deriving eq. (A.18) we used
\be\label{AA.18}
\left[\frac{\delta}{\delta\bar \xi^y}-g D_\mu(\bar
A)^{yz}\frac{\delta}{\delta\bar\beta^z_\mu}\right]
\Gamma_k[A,\bar A,\xi,\bar \xi;\bar\beta,\bar\gamma]=0\ee
which follows from the equation of motion of the antighost.

Equation (\ref{AA.16}) is the generating
relation for the modified Ward identities
which we wanted to derive. In conventional
Yang-Mills theory, without IR-cutoff,
the r.h.s. of (\ref{AA.16}) is zero. The traces on the
r.h.s. of (\ref{AA.16}) lead to a violation
of the usual Ward identities for nonvanishing
values of $k$. As $k$ approaches zero,
$R_k$ and hence $\Delta^{\rm(BRS)}_k$ vanishes
and we recover the conventional Ward-Takahashi
identities. Eq. (\ref{AA.16}) is equivalent
to a similar identity derived in ref. \cite{Ell2}
using  a different gauge.
\vspace{3mm}

The modified Ward identities (\ref{AA.16}) are not the only conditions
which the average action $\Gamma_k$ has to satisfy. Next we derive
a relation which constrains its dependence on the background gauge
field $\bar A_\mu$. Taking the
$\bar A$-derivative of (\ref{AA.1}) one obtains
\ba\label{AA.19}
&&-\frac{\delta W_k}{\delta \bar A_\mu^z(y)}
=-(R_k\bar a)^z_\mu(y)+\frac{1}{\alpha}(\bar D\otimes
\bar D \bar a)^z_\mu(y)+K^z_\mu(y)\nonumber\\
&&+\int d^dx\left\lbrace\frac{1}{2}\bar a_\nu\frac{\delta(R_k-
\frac{1}{\alpha}\bar D\otimes\bar D)_{\nu\tau}}{\delta\bar A
_\mu^z(y)}\bar a_\tau+\bar \xi\frac{\delta R_k}
{\delta\bar A_\mu^z(y)}\xi
\right\rbrace\nonumber\\
&&+\frac{1}{2}\Tr_{xcL}\left[\frac{\delta^2W_k}
{\delta K_\tau\delta K_\nu}
\frac{\delta}{\delta\bar A_\mu^z(y)}(R_k-\frac{1}{\alpha}
\bar D\otimes\bar D)_{\nu\tau}\right]\nonumber\\
&&-\Tr_{xc}\left[\frac{\delta^2W_k}{\delta\bar\sigma\delta\sigma}
\frac{\delta}{\delta\bar A^z_\mu(y)}R_k\right]\nonumber\\
&&-g^2\bar \xi^v(y)f^{vzw}\frac{\delta W_k}{\delta\bar\beta^w_\mu(y)}
+g^2f^{zvw}\frac{\delta^2W_k}{\delta\sigma^v(y)
\delta\bar\beta^w_\mu(y)}
\ea
with $(\bar D\otimes\bar D)^{vw}_{\mu\nu}\equiv D_\mu(\bar A)^{vz}
D_\nu(\bar A)^{zw}$. Upon Legendre transforming eq. (\ref{AA.19})
and switching from $\tilde\Gamma_k$ to $\Gamma_k$ one arrives at the
following result
\ba\label{AA.20}
&&\frac{\delta}{\delta\bar A^z_\mu(y)}\Gamma_k'[A,\bar A,\xi,\bar
\xi;\bar\beta,\bar\gamma]
=-g^2\bar \xi^v(y)f^{zvw}\frac{\delta\Gamma_k}
{\delta\bar\beta^w_\mu(y)}
\nonumber\\
&&+\frac{1}{2}\Tr_{xcL}\left[\left(\Gamma_k^{(2)}+R_k\right)^{-1}
_{AA}\frac{\delta}{\delta\bar A_\mu^z(y)}\left(R_k-\frac{1}{\alpha}
\bar D\otimes\bar D\right)_{AA}\right]\nonumber\\
&&-\Tr_{xc}\left[\left(\Gamma_k^{(2)}+R_k\right)^{-1}_{\bar\xi\xi}
\frac{\delta R_{k\bar\xi\xi}}{\delta\bar A^z_\mu(y)}\right]\nonumber\\
&&+g^2\int d^dx\ tr_c\left[T^z\left(\Gamma_k^{(2)}+R_k\right)^{-1}
_{\bar\xi(y)\varphi(x)}\frac{\delta^2\Gamma_k}{\delta\varphi(x)
\delta\bar\beta_\mu(y)}\right]\ea
Again, $\varphi\equiv(A,\xi,\bar \xi)$ is summed over and $(T^z)^{yw}=
-if^{zyw}$. Note that the r.h.s. of eq. (\ref{AA.20}) does not
vanish even for $k\to0$. The $\bar D\otimes\bar D$-piece of the
2nd term and the 4th term on the r.h.s. of (\ref{AA.20}) survive
this limit.
\vspace{3mm}

So far we were deriving general identities which constrain the
form of the exact functional $\Gamma_k$. Let us now
ask what they imply if we truncate the space of actions. In the
present paper we neglect the $k$-evolution of the ghost sector by
making an ansatz which keeps the classical form of the corresponding
terms in the action:
\be\label{A.23}
\Gamma_k[A,\bar A,\xi, \bar\xi;\bar\beta,\bar\gamma]
=\Gamma_k[A,\bar A]+\Gamma_{\rm gh}\ee
\ba\label{A.24}
\Gamma_{\rm gh}&=&-\int d^dx\ \bar\xi D_\mu(\bar A)D_\mu(A)\xi\nonumber\\
&&-\int d^dx\left\{\frac{1}{g}\bar\beta^v_\mu(D_\mu(A)^{vw}
\xi^w+\frac{1}{2}\bar\gamma^u f^{uvw}\xi^v\xi^w\right\}\ea
If we insert this truncation into the general evolution equation
(\ref{AA.13}), we obtain precisely eq. (\ref{1.1}), which is our
starting point in the main body of the paper. A generic functional
$\Gamma_k[A,\bar A]$ can be decomposed according to
\be\label{A.25}
\Gamma_k[A,\bar A]=\bar\Gamma_k[A]+\frac{1}{2\alpha}\int d^dx
[D_\mu(\bar A)(A_\mu-\bar A_\mu)]^2+\hat\Gamma_k^{\rm gauge}[A,\bar A]
\ee
where $\bar\Gamma_k$ is defined by equating the two-gauge fields:
$\bar\Gamma_k[A]\equiv \Gamma_k[A,A]$. The remainder $\Gamma_k[A,\bar
A]-\bar\Gamma_k[A]$ is further decomposed in the classical
gauge-fixing term plus a correction to it, $\hat\Gamma_k^{\rm gauge}$,
which also contains the counterterms.
Note that $\hat\Gamma_k^{\rm gauge}[A,A]=0$ for equal gauge
and background fields. We observe that
$\bar\Gamma_k
[A]$ is a gauge-invariant functional of $A_\mu$ and $\Gamma_k[A,\bar A]$
is invariant under a simultaneous gauge transformation of $A$
and $\bar A$.

In the present paper we make the further approximation of neglecting
quantum corrections to the gauge fixing term
and $k$-dependent counterterms by setting
$\hat\Gamma_k^{\rm gauge}=0$. Then (\ref{A.25}) coincides with
eq. (\ref{2.1}) for
\be\label{A.26}
\bar\Gamma_k[A]=\int d^dx\ W_k(\frac{1}{4}F^2).\ee
The important question is whether this truncation is consistent
with the Ward-Takahashi identities (\ref{AA.16}) and the
$\bar A$-derivative (\ref{AA.20}), respectively. If we insert
(\ref{A.23})-(\ref{A.25}) into (\ref{AA.16}),
for instance, we find that $\bar\Gamma_k$ drops out from the l.h.s.
of this equation. We are left with a condition for $\hat\Gamma_k
^{\rm gauge}$
\be\label{A.27}
-\frac{1}{g}\int d^dx\frac{\delta\hat\Gamma^{\rm gauge}}{\delta\bar
A_\mu^z(x)}(D_\mu(A)\xi)^z(x)=\Delta_k^{(\rm BRS)}\ee
The term $\Delta^{(\rm BRS)}_k$ (\ref{AA.16a})
vanishes for $k\to0$ but is non-zero for $k>0$. Our
approximation $\hat\Gamma_k^{\rm gauge}\equiv0$ is consistent
provided these terms can be neglected. We note that the traces
appearing in (\ref{A.27}) are related to higher loop effects.
Beyond a loop approximation our neglection of $\hat\Gamma^{\rm gauge}$
is a non-trivial assumption. We emphasize that because of its
gauge invariance the functional $\bar\Gamma_k[A]$ does not appear
on the l.h.s. of the Ward identities. Therefore the Ward identities
do not imply any further condition for $\bar\Gamma_k$. This means
that, {\it within the approximations made, we may write down any
ansatz for $\bar\Gamma_k$ as long as it is gauge-invariant}.

Similar remarks apply to the identity for the $\bar A$-dependence
(\ref{AA.20}). If we insert (\ref{A.23}) with (\ref{A.25}) into (\ref
{AA.20}) the first term on the r.h.s. is cancelled by $\delta\Gamma
_{\rm gh}/\delta\bar A_\mu^z(y)$. One obtains
\ba\label{A.28}
\frac{\delta\hat\Gamma_k^{\rm gauge}}{\delta\bar A_\mu^z(y)}&=&-
\frac{1}{2\alpha}\ Tr_{xcL}\left[\left(\Gamma_k^{(2)}+
R_k\right)^{-1}_{AA}
\frac{\delta\bar D\otimes\bar D}{\delta\bar A_\mu^z(y)}\right]
\nonumber\\
&&+g^2\int d^dx\ tr_c\left[T^z\left(\Gamma_k^{(2)}+R_k\right)^{-1}_
{\bar\xi(y)\varphi(x)}\frac{\delta^2\Gamma_k}
{\delta\varphi(x)\delta\bar\beta_\mu(y)}\right]\nonumber\\
&&+Tr[R_k(...)]\ea
Consistency of the truncation $\hat\Gamma_k^{\rm gauge}=0$
requires that we neglect the traces on the r.h.s. of (\ref{A.28}).
Contrary to the case of the Ward identities, not all of these terms
vanish obviously for $k\to0$. We observe again that, within the
present approximation, the equation for the $\bar A$-dependence
does not impose any restriction on $\bar\Gamma_k$.

We finally observe that the flow equation (\ref{1.1}) can be
rewritten in close analogy to a one-loop formula:
\ba\label{1.4}
\frac{\partial}{\partial t} \Gamma_k[A,\bar
A]&=&\frac{1}{2}\frac{D}{Dt}\Tr_{xcL}\ln\left[\Gamma^{(2)}_k[A,\bar
A]+R_k\left(\Gamma^{(2)}_k[\bar A,\bar A]\right)\right]\nonumber\\
&&-\frac{D}{Dt}\Tr_{xc}\ln\left[-D^\mu[A]D_\mu[\bar A]+R_k(-D^2[\bar
A])\right]
\ea
The derivative $\frac{D}{Dt}$ acts only on the explicit $k$-dependence of
the function $R_k$, but not on $\Gamma^{(2)}_k [A,\bar A]$.
It is now
easy to describe the relation between the effective average action
$\Gamma_k$ and the conventional perturbative
effective action. Let us first briefly discuss the
approximation $\frac{D}{Dt}\to\frac{\partial}{\partial t}$ in eq.
(\ref{1.3}). This amounts to neglecting the running of $\Gamma_k$ on the
r.h.s. of the evolution equation. It is then trivial to solve it
explicitly:
\ba\label{1.5}
\Gamma_k[A,\bar A]&=&\Gamma_\Lambda[A,\bar
A]+\frac{1}{2}\Tr_{xcL}\left\{\ln\left[\Gamma_k^{(2)}[A,\bar
A]+R_k(\Gamma_k^{(2)}[\bar A,\bar A])\right]\right.\nonumber\\
&&\left.-\ln\left[\Gamma_\Lambda^{(2)}[A,\bar A]+R_\Lambda
(\Gamma_\Lambda^{(2)}[\bar A,\bar A])\right]\right\}\nonumber\\
&&-\Tr_{xc}\left\{\ln\left[-D^\mu[A]D_\mu[\bar A]+R_k(-D^2[\bar
A])\right]\right.\nonumber\\
&&\left.-\ln\left[-D^\mu[A]D_\mu[\bar A]+R_\Lambda(-D^2[\bar
A])\right]\right\}
+O\left(\frac{\partial}{\partial t}\Gamma^{(2)}_k\right)
\ea
with $\Lambda$ some appropriate high momentum scale (ultraviolet cutoff)
where we may identify $\Gamma_\Lambda$ with the classical action $S$
including a gauge fixing term and counterterms.
This formula has  a similar structure as a
regularized expression for the conventional one-loop effective action in
the background gauge \cite{abb,dit}. There are two important differences,
however:

(i) The second variation of the classical action, $S^{(2)}$, is
replaced by $\Gamma^{(2)}_k$. This implements a kind of ``renormalization
group improvement'' and transforms (\ref{1.4}) into a sort of ``gap
equation''.

(ii)  The effective average action contains an explicit infrared cutoff
$R_k$. For the choice (\ref{1.2}) one has
\be\label{1.6}
\lim_{u\to\infty} R_k(u)=0,\qquad \lim_{u\to 0} R_k(u)=Z_k k^2.
\ee
Effectively,
a $k$-dependent mass-type term is added to the inverse propagator
$\Gamma^{(2)}_k$ for the low frequency modes $(u\to 0)$,
but it is absent for
the high frequency modes $(u\to \infty)$.
Despite the similarity of (\ref{1.3}) with a one-loop expression, we
stress that, for $k\to 0$, the solution of the original renormalization
group equation where $D/Dt$ does not act on $\Gamma_k^{(2)}$
equals  the {\it exact}
effective action which includes contributions  from all orders of the loop
expansion\footnote{This holds for the exact flow equation (\ref{AA.13}),
whereas (1.1) involves already an approximation in the ghost sector.}.

For a detailed discussion of the approximation (\ref{1.4}) in
the case of the abelian Higgs model we refer to \cite{wet}, and to ref.
\cite{rwe} for the corresponding  nonperturbative evolution equations
of this model.

\section*{Appendix B}
\renewcommand{\theequation}{B.\arabic{equation}}
\setcounter{equation}{0}

In this appendix we discuss the technical steps needed for the
derivation of the truncated flow equation (\ref{3.4}) from
the exact equation (\ref{AA.13}). Upon
performing the second variation of the ansatz (\ref{2.1}),
\be\label{2.2}
\delta^2\Gamma_k[A,\bar A]=\int d^dx\delta A^\mu_y\Gamma_k^{(2)}[A,\bar A]
^{yz}_{\mu\nu}\delta A^\nu_z\ee
we arrive at the following operator $\Gamma_k^{(2)}$:
\be\label{2.3}
\Gamma_k^{(2)}[A,\bar A]^{yz}_{\mu\nu}=W_k'(\theta)({\cal D}_T[A]-{\cal D}
_L[A])^{yz}_{\mu\nu}
+W_k''(\theta){\cal S}^{yz}_{\mu\nu}[A]+\frac{1}{\alpha_k}({\cal D}_L[\bar
A])^{yz}
_{\mu\nu}\ee
with
\be\label{2.4}
\theta=\frac{1}{4}F^z_{\mu\nu}F^{\mu\nu}_z\ee
Here we used the notation ($w, y, z$ are adjoint group indices and $\bar
g$ is the (bare) gauge coupling)
\ba\label{2.5}
({\cal D}_T)^{yz}_{\mu\nu}&=&(-D^2\delta_{\mu\nu}+2i\bar g
F_{\mu\nu})^{yz}
\nonumber\\
({\cal D}_L)^{yz}_{\mu\nu}&=&-(D\otimes D)^{yz}_{\mu\nu}=-D^{yw}_\mu
D^{wz}_\nu\nonumber\\
{\cal S}^{yz}_{\mu\nu}&=&F^y_{\mu\rho}F^w_{\sigma\nu}(D^\rho
D^\sigma)^{wz}
\ea
with the covariant derivative $(D_\mu[A])^{yw}=\partial_\mu\delta^{yw}
-i\bar g A^z_\mu(T_z)^{yw}$ in the adjoint representation and
$F^{yw}_{\mu\nu}=F^z_{\mu\nu}(T_z)^{yw}$.
Moreover, $W_k'$
and $W_k''$ denote the first and the second derivative of $W_k$ with
respect
to $\theta$.

In writing down eq. (\ref{2.5}) we made the additional
assumption that the field strength $F_{\mu\nu}[A]$ is covariantly
constant $F^z_{\mu\nu;\rho}=0$ or
\be\label{2.6}
[D_\rho[A],F_{\mu\nu}[A]]=0\ee
It is easy to see that (\ref{2.7}) and (\ref{2.8}) obey the
condition (\ref{2.6}). The choice (\ref{2.7}) has the advantage that it
allows for an explicit diagonalization of the operator $\Gamma_k^{(2)}$.

We note that $W_k$ can be extracted from $\Gamma_k[A,A]$ which is a
gauge-invariant functional of $A$ obtained by putting $\bar A=A$. It is
therefore sufficient to know $\Gamma_k^{(2)}[A,A]$.

Before turning to the evolution equation, we list a few special
properties
of the covariantly constant fields, which will prove helpful later on.
{}From (\ref{2.7})
it follows that $A_\mu$ satisfies the classical Yang-Mills equations
$D^\mu
F_{\mu\nu}=0$. This in turn is sufficient to prove that the operators
${\cal D}_T$ and ${\cal D}_L$ commute. As a consequence, one may define
generalized projection operators \cite{reu}
\ba\label{2.9}
P_L&=&{\cal D}^{-1}_T{\cal D}_L\nonumber\\
P_T&=&1-P_L\ea
which satisfy $P_{T,L}^2=P_{T,L},P_T+P_L=1$ and $P_TP_L=0=P_LP_T$. For
$A_\mu=0$ they reduce to the standard projectors on transverse and
longitudinal modes:
\ba\label{2.10}
(P_T^{(0)})_{\mu\nu}&=&\delta_{\mu\nu}-\partial_\mu\partial_\nu/\partial^2
\nonumber\\
(P_L^{(0)})_{\mu\nu}&=&\partial_\mu\partial_\nu/\partial^2\ea
Furthermore, if $A_\mu$ is of the form (\ref{2.7}), it is natural
to define another pair of orthogonal projectors,
\be\label{2.11}
P_\perp^{yz}=\delta^{yz}-n^yn^z,\quad P_{\parallel}^{yz}=n^yn^z,\ee
which project on the spaces perpendicular and parallel to $n^z$,
respectively. For the vector potential (\ref{2.7}) the matrix
$A^w_\mu T_w$ reads in the adjoint representation
\be\label{2.12}
A^{yz}_\mu(x)\equiv(A^w_\mu T_w)^{yz}=if^{ywz}n_w{\sf A}_\mu(x)\ee
The antisymmetry of the structure constants $f^{ywz}$ implies that
$P_{\parallel}$ and $P_\perp$ commute with $D_\mu,D^2,{\cal D}_T, {\cal D}
_L$ and $F_{\mu\nu}$, and that
\be\label{2.13}
P_{\parallel}A_\mu=0,\quad P_\parallel D_\mu=P_\parallel
\partial_\mu,\quad
P_\parallel({\cal D}_T)_{\mu\nu}=-\partial^2\delta_{\mu\nu}P_\parallel\ee
The operator ${\cal S}$ from (\ref{2.5}) factorizes according to
\ba\label{2.14}
{\cal S}^{yz}_{\mu\nu}&=&P^{yz}_{||}s_{\mu\nu},\nonumber\\
s_{\mu\nu}&=&{\sf F}_{\mu\rho}{\sf
F}_{\sigma\nu}\partial^\rho\partial^\sigma
\ea
Hence ${\cal S}$ commutes with $D^2, {\cal D}_L$ and ${\cal D}_T$ because
it annihilates the gauge-field
contained in these operators:
\be\label{2.15}
{\cal S}D^2={\cal S}\partial^2,\quad {\cal S}{\cal D}_T=-{\cal
S}\partial^2,
\quad {\cal S} D\otimes D={\cal S}\partial\otimes\partial\ee

In physical terms this means that those components of the gauge
fluctuations $\delta A^z_\mu\equiv a^z_\mu$ which are parallel
to $n^z$ decouple from $A_\mu$ to some extent. In fact, in terms of the
projections $a_\mu^{\perp,||}=P_{\perp,||}a_\mu$ the quadratic action
(\ref{2.2})
with (\ref{2.6}) reads
\ba\label{2.16}
&&\delta^2\Gamma_k[A,A]=\int d^dx\left\lbrace a^{||\mu}_z[-\partial^2
W_k'\delta_{\mu\nu}+(W_k'-\frac{1}{\alpha_k})\partial_\mu\partial_\nu
\right.
+W_k''s_{\mu\nu}]a^{||\nu z}\nonumber\\
&&\left.+a^{\perp \mu}_y[W_k'{\cal D}_T+(\frac{1}{\alpha_k}-W_k'){\cal
D}_L]^{yz}_{\mu\nu}a^{\perp\nu}_z\right\rbrace\ea
We observe that the $a^{||}$-modes couple to the external field only via
the
derivatives of $W_k\equiv W_k(\frac{1}{2}B^2)$. In a conventional
one-loop calculation one uses the classical Yang-Mills Lagrangian
$\frac{1}{4}F^2_{\mu\nu}$ rather than $W_k(\frac{1}{4} F^2_{\mu\nu})$.
In that case the quadratic action for the small fluctuations is given by
(\ref{2.16}) with $W_k'=1$ and $W_k''=0$. Hence the one-loop determinant
resulting from the integration over $a^{||}$ is field-independent and may
be ignored. In the present case, the $a^{||}$-modes are important for the
``renormalization group improvement'', however.

The quadratic form
(\ref{2.16}) can be diagonalized even further by introducing the
longitudinal and transversal projections
\ba\label{2.17}
&&a^{||,L}=P_La^{||}=P_L^{(0)}a^{||},\quad
a^{||,T}=P_Ta^{||}=P_T^{(0)}a^{||}
\nonumber\\
&&a^{\perp,L}=P_L a^\perp,\quad a^{\perp,T}=P_Ta^\perp\ea
By virtue of $P_L^{(0)}s=sP_L^{(0)}=0,P_T^{(0)}s=sP_T^{(0)}=s$
and $[P_{L(T)},\theta]=0$ one obtains
\ba\label{2.18}
\delta^2\Gamma_k[A,A]&=&\int d^dx\left\lbrace a^{||,T,\mu}_z[-\partial^2
W_k'\delta_{\mu\nu}
+W_k''s_{\mu\nu}]a^{||,T,\nu,z}\right.\nonumber\\
&&+\frac{1}{\alpha_k}a^{||,L,\mu}_z[-\partial^2]a^{||,L,z}_\mu\nonumber\\
&&+a^{\perp,T,\mu}_y[W_k'{\cal
D}_T]^{yz}_{\mu\nu}a^{\perp,T,\nu}_z\nonumber\\
&&\left.+\frac{1}{\alpha_k}a^{\perp,L,\mu}_y[{\cal
D}_T]^{yz}_{\mu\nu}a^{\perp,L,\nu}_z\right\rbrace\ea
This block-diagonal form of $\Gamma_k^{(2)}$ will facilitate the evolution
of the traces occurring in the evaluation equation. For example,
$a^{||,L}$
gives no $A$-dependent contribution and, except for an irrelevant
constant,
the only dependence of $\Gamma_k$ on $\alpha_k$ arises from $a^{\perp,L}$.
Writing
\be\label{2.19}
\Gamma^{(2)}[A,A]=\Gamma_1^{(2)}+\Gamma_2^{(2)}+\Gamma_3^{(2)}+
\Gamma_4^{(2)}\ee
where, in an obvious notation
\ba\label{2.20}
&&\Gamma_1^{(2)}=P_{||}P_T\Gamma^{(2)}_{||,T}P_{||}P_T,\quad
\Gamma_2^{(2)}=P_{||}P_L\Gamma^{(2)}_{||,L}P_{||}P_L\nonumber\\
&&\Gamma_3^{(2)}=P_{\perp}P_T\Gamma^{(2)}_{\perp,T}P_{\perp}P_T,
\quad\Gamma_4^{(2)}=P_{\perp}P_L\Gamma^{(2)}_{\perp,L}P_{\perp}P_L\ea
with
$[P_{||,\perp},P_{L,T}]=0$ and
\ba\label{2.21}
&&\Gamma_A^{(2)}\Gamma_B^{(2)}=0\quad{\rm for}\quad A\not=B\nonumber\\
&&[\Gamma_A^{(2)},\Gamma_B^{(2)}]=0\ea
one obtains ([$\Gamma^{(2)}_{||,T},P_{||}P_T]=0$ etc.)
\ba\label{2.22}
R_k(\Gamma^{(2)})&=&P_{||}P_TR_k(\Gamma_{||,T}^{(2)})P_{||}P_T
+P_{||}P_LR_k(\Gamma_{||,L}^{(2)})P_{||}P_L\nonumber\\
&&+P_{\perp}P_TR_k(\Gamma_{\perp,T}^{(2)})P_{\perp}P_T
+P_{\perp}P_LR_k(\Gamma_{\perp,L}^{(2)})P_{\perp}P_L\ea

We will choose the matrix
${\cal Z}_k$ in the
definition of $R_k$ (\ref{1.2}) as ${\cal Z}_k=1$ for the ghosts and
\be\label{3.1}
{\cal Z}_k=Z_kP_T[\bar A]+\tilde Z_kP_L[\bar A]\ee
for the gauge boson degrees of freedom. Here $Z_k,\tilde Z_k$ are
$k$-dependent constants and we observe that the choice (\ref{3.1}) is
compatible with (\ref{2.22}). If we insert the truncation (\ref{2.1}) into
(\ref{1.1}) with $\bar A=A$, we obtain $(\theta=\frac{1}{2}B^2)$
\ba\label{3.2}
\Omega\frac{\partial}{\partial t}W_k(\theta)&=&\frac{1}{2}{\rm
Tr}_{xcL}[H(
\Gamma_k^{(2)}[A,A])]\nonumber\\
&&-{\rm Tr}_{xc}[H_G(-D^2[A])]\nonumber\\
&&+\frac{1}{2}{\rm Tr}_{xcL}[P_\perp P_L(\tilde H(\Gamma_{\perp
L}^{(2)})-H(\Gamma_{\perp
L}^{(2)}))]\ea
where $\Gamma_k^{(2)}$ is given by (\ref{2.6}), $H(u)$ is defined
by eq. (\ref{3.3})
and $\Omega\equiv \int d^dx$.
\vspace{4mm}

Let us pause here for a moment and derive a set of trace
identities which will be needed for the evaluation of
(\ref{3.2}). For the
covariantly constant background (\ref{2.7}), eq.
(\ref{2.13}) implies for any function $f$
\be\label{A.1}
\Tr_{xcL}[
P_{\parallel}f(D_\mu,P_{\parallel},P_\perp)]=\Tr_{xL}[f(\partial_\mu,1,0)]
\ee
because $\Tr_c[P_{\parallel}]=n^zn_z=1.$ Writing $P_\perp=1-P_{\parallel}$
and exploiting $ {\cal S}\propto P_{\parallel}$ it is also easy to see
that
\be\label{A.2}
\Tr_{xcL}[P_\perp f(D_\mu,{\cal
S})]=\Tr_{xcL}[f(D_\mu,0)]-\Tr_{xL}[f(\partial_\mu,0)].
\ee
Since  ${\sf F}_{\mu\nu}$ is a constant matrix, the operator $s_{\mu\nu}$
of (\ref{2.14}) commutes with $P_L^{(0)}$ and $P^{(0)}_T$ and satisfies
$\partial^\mu s_{\mu\nu}=0$. This fact can be used to show that
\ba\label{A.3}
&&\Tr_{xL}[f(P_L^{(0)},P^{(0)}_T;s)]\nonumber\\
&=&\Tr_{xL}[f(0,1;s)]+\Tr_x[f(1,0;0)]-\Tr_x[f(0,1;0)]\ea
If one subtracts the same expression with $s=0$ one obtains
\ba\label{A.4}
&&\Tr_{xL}[f(P_L^{(0)},P_T^{(0)};s)-f(P_L^{(0)},P_T^{(0)};0)]\nonumber\\
&=&\Tr_{xL}[f(0,1;s)]-d\Tr_x[f(0,1;0)]
\ea
In the last step we used that $\Tr_{xL}=d\Tr_x$ for an operator
$\sim\delta_{\mu\nu}$. In the above identities the function $f$ may also
depend on further operators provided they commute with those displayed
explicitly and do not introduce any additional colour or Lorentz index
structures.

For the evaluation of $U_1$ in eq. (\ref{B.32}) we need another important
relation:
\be\label{A.5}
\Tr_{xcL}[P_L f ({\cal D}_T)]=\Tr_{xc}[f(-D^2)]
\ee
It follows from the fact that the operator $({\cal
D}_T)_{\mu\nu}=-D^2\delta_{\mu\nu}+2i\bar g F_{\mu\nu}$, when restricted
to the space of longitudinal modes $(a_\mu=(P_L)^\nu_\mu a_\nu)$, has the
same spectrum as $-D^2$ acting on Lorentz scalars. The proof makes
essential use of the identity
\be\label{A.6}
{\cal D}_T(D\otimes D)=(D\otimes D){\cal D}_T=-(D\otimes D)(D\otimes D)
\ee
which holds true whenever the gauge field contained in the covariant
derivatives obeys $D^\mu F_{\mu\nu}=0$.
\vspace{4mm}

Equipped with the above trace identities, we now resume the evaluation
of the flow equation (\ref{3.2}).
The first trace on the
r.h.s. of eq. (\ref{3.2}) can be simplified as follows.
Inserting a factor of  $1=P_{\parallel}+P_\perp$ leads to the
decomposition
\be\label{B.28}
\Tr_{xcL}\left[H(\Gamma^{(2)}_k[A,A])\right]=T^{\parallel}_1+T^\perp_1
\ee
with
\ba\label{B.29}
T^{\pl}_1&=&\Tr_{xcL}\left[P_\pl H(W'_k{\cal
D}_T+[W'_k-\frac{1}{\alpha_k}]D\otimes D+W''_k{\cal S})\right]\nonumber\\
&=&\Tr_{xL}\left[H(-W'_k\partial^2+[W'_k-
\frac{1}{\alpha_k}]\partial\otimes
\partial+W''_k s\right]
\ea
where (\ref{A.1}) was used, and
\ba\label{B.30}
T_1^\perp&=&\Tr_{xcL}\left[P_\perp H(W'_k{\cal
D}_T+[W'_k-\frac{1}{\alpha_k}]D\otimes D+ W''_k{\cal S})\right]\nonumber\\
&=& \Tr_{xcL}\left[ H(W'_k{\cal D}_T+[W'_k-\frac{1}{\alpha_k}] D\otimes
D)\right]\nonumber\\
&&-\Tr_{xL}\left[H(-W'_k\partial^2
+[W'_k-\frac{1}{\alpha_k}]\partial\otimes\partial)\right]
\ea
where (\ref{A.2}) was exploited. Let us write
\be\label{B.31}
\Tr_{xcL}\left[H(\Gamma^{(2)}_k[A,A])\right] =U_1+U_2
\ee
with $U_1$ the ``nonabelian'' trace
\be\label{B.32}
U_1=\Tr_{xcL}\left[ H(W'_k{\cal D}_T+[W'_k-\frac{1}{\alpha_k}]D\otimes
D)\right]\ee
and $U_2$ the sum of $T_1^{\pl}$ and the second term of (\ref{B.30}):
\ba\label{B.33}
U_2&=&\Tr_{xL}\left[H\left(-\partial^2[W_k'
P^{(0)}_T+\frac{1}{\alpha_k}P_L^{(0)}]+W_k'' s\right)\right]\nonumber\\
&&-\Tr_{xL}\left[H\left(-\partial^2[W_k' P^{(0)}_T +\frac{1}{\alpha_k}
P_L^{(0)}]\right)\right]\ea
It is quite remarkable that if we now apply the identity (\ref{A.4}) to
$U_2$, the longitudinal contribution drops out completely and the result
becomes independent of the gauge fixing parameter $\alpha_k$:
\ba\label{B.34}
U_2=\Tr_{xL}[H(-\partial^2 W_k'+W_k'' s)]
-d\Tr_x[H(-\partial^2 W_k')]\ea
The operators entering (\ref{B.34}) are easily diagonalized in a plane-wave
basis. A standard calculation yields, for $W_k'>0,W_k'+B^2W_k''>0$,
\be\label{B.35}
\Omega^{-1}U_2=2v_d\left(\frac{1}{W_k'+B^2
W_k''}-\frac{1}{W_k'}\right)\left(\frac{1}
{W_k'}\right)^{\frac{d}{2}-1}\int^\infty_0 dx \ x^{\frac{d}{2}-1}H(x)\ee
with
$v_d=[2^{d+1}\pi^{d/2}\Gamma(d/2)]^{-1}$. (As always, the argument of
$W_k$ and its derivatives is understood to be $\frac{1}{2} B^2$.)

Next  let us simplify the trace $U_1$ by inserting a pair of projectors:
\ba\label{B.36}
U_1&=&\Tr_{xcL}\left[ P_T H({\cal D}_T[W'_kP_T+\frac{1}{\alpha_k}
P_L])\right]
+\Tr_{xcL}\left[ P_L H ({\cal D}_T[W'_k P_T+\frac{1}{\alpha_k}
P_L])\right]\nonumber\\
&=&\Tr_{xcL}[P_T H(W'_k{\cal D}_T)]+\Tr_{xcL}[P_L
H(\frac{1}{\alpha_k}{\cal D}_T)]\nonumber\\
&=&\Tr_{xcL}[H(W'_k{\cal D}_T)]+\triangle U_1(\alpha_k).\ea
The $\alpha$-dependence of $U_1$ is contained in
\ba\label{B.37}
\triangle U_1(\alpha_k)&=&\Tr_{xcL}\left[ P_L\{ H(\frac{1}{\alpha_k}{\cal
D}_T)-H(W'_k{\cal D}_T)\}\right]\nonumber\\
&=&\Tr_{xc}\left[H(-\frac{1}{\alpha_k} D^2)-H(-W'_kD^2)\right].\ea
In the last line of (\ref{B.37}) we made use of the identity (\ref{A.5}).

By a similar combination of the trace identities we can also evaluate the
last term on the r.h.s. of (\ref{3.2})
\ba\label{B.38}
&&\Tr_{xcL}\left\{P_\perp P_L\left(\tilde H\left(\frac{{\cal
D}_T}{\alpha_k}\right)-H\left(\frac{{\cal
D}_T}{\alpha_k}\right)\right)\right\}\nonumber\\
&=&\Tr_{xc}\left\{\tilde
H\left(-\frac{D^2}{\alpha_k}\right)-H\left(-\frac{D^2}
{\alpha_k}\right)\right\}\nonumber\\
&&-\Tr_x\left\{\tilde
H\left(-\frac{\partial^2}{\alpha_k}\right)-H
\left(-\frac{\partial^2}{\alpha_k}\right)\right\}.\ea
Evaluating the second term in a plane wave basis yields
\be\label{B.39}
\frac{1}{2\Omega}\Tr_x\left\{\tilde
H\left(-\frac{\partial^2}{\alpha_k}\right)-H
\left(-\frac{\partial^2}{\alpha_k}\right)\right\}=v_d\int^\infty_0 dx\
x^{\frac{d}{2}-1}\left(\tilde
H\left(\frac{x}{\alpha_k}\right)-H\left(\frac{x}{\alpha_k}
\right)\right)\ee
At this point we have exploited the various trace identities  as much as
possible.
Combining these results yields the flow equation (\ref{3.4}).

\section*{Appendix C}

\renewcommand{\theequation}{C.\arabic{equation}}
\setcounter{equation}{0}

In this appendix we discuss the group theoretical factors
$\sum_\ell\nu_\ell^{2m}$ appearing in the Euler-McLaurin expansion
of the spectral sums.

The LHS of the evolution equation is
$\partial_t W_k$. The argument of $W_k$ is $\frac{1}{4} F^z_{\mu\nu}
F_z^{\mu\nu}=\frac{1}{2} B^2$ which is  manifestly independent of the unit
vector $n^z$ which specifies the direction of the field in ``color
space''. The r.h.s. of the evolution equation  consists of expansions such
as (\ref{3.9}) which involve the factors $\sum_\ell\nu^{2m}_\ell$. As
$\{\nu_\ell\}$ are the eigenvalues of $n^zT_z$, we can rewrite them as
\be\label{C.1}
\sum_\ell \nu_\ell^{2m}=n^{z_1} n^{z_2}\cdots n^{z_{2m}} \Tr_c[T_{z_1}
T_{z_2}\cdots T_{z_{2m}}]\ee
where the trace is in the adjoint representation. The question is whether
the invariants (\ref{C.1}) are all independent of the direction of $n^z$.
In appendix D we explain in detail that generically this is \underbar{not}
the case. If the orbit space  of the gauge group in the adjoint
representation is nontrivial, different $n$'s  can lead to different sums
$\sum_\ell \nu^{2m}_\ell$. The resolution to this puzzle is as follows.
For $m=1$ we can use the standard orthogonality relation
\be\label{C.2}
\Tr_c[T_y T_z]=N\delta_{yz}\ee
to prove that $\sum_\ell\nu^2_\ell=N n^z n_z=N$ is independent of the
direction of $n$. Likewise, if the symmetric invariant tensor $\Tr_c\left[
T_{(z_1}\cdots T_{z_{2m})}\right]$ is proportional to the trivial one,
$\delta_{(z_1z_2}\delta_{z_3z_4}\cdots\delta_{z_{2m-1}z_{2m})}$, we can
again use the normalization condition $n^zn_z=1$ to show that the r.h.s.
of (\ref{C.1}) is independent of $n$.

The situation changes if there
exists a totally symmetric invariant tensor ${\cal T}_{z_1z_2\cdots
z_{2m}}$ which is different from the trivial one. Then we might have
\be\label{C.3}
\Tr_c\left[ T_{(z_1}\cdots T_{z_{2m})}\right]=\tau_m\delta_{(z_1z_2}\cdots
\delta_{z_{2m-1}z_{2m})}+{\cal T}_{z_1z_2\cdots z_{2m}}\ee
with some coefficient $\tau_m$. (If there exists more than one ${\cal T}$
an appropriate sum is implied.) In general $n^{z_1} n^{z_2}...{\cal
T}_{z_1z_2\cdots}$ will be direction dependent \cite{rai}.
If some invariant tensor ${\cal T}$ exists, the correct way of deriving
the evolution equation for $W_k$ is to compare coefficients of a fixed
tensor structure on both sides of the equation. Clearly the l.h.s.,
$\partial
_tW_k(\frac{1}{4}F^2_{\mu\nu})$, gives rise to the trivial tensor
structure
only. Therefore only the $\tau_m$-piece of (\ref{C.3}) should be
kept in (\ref{C.1}) and the ($n$-dependent) part coming from ${\cal T}$
has to be discarded. Thus (\ref{3.9}) may be used on the r.h.s. of the
equation for $W_k$ provided we interpret $\sum \nu^{2m}_l$ as the
coefficient $\tau_m$.

On the other hand, a nontrivial ${\cal T}$ permits us
to construct additional invariants from an even number of covariantly
conserved $F_{\mu\nu}$. Then the truncation $W(\theta)$ is not sufficient
any more to parametrize the most general effective action for constant
magnetic fields of the type introduced in sect. 2 (with covariantly
constant $F_{\mu\nu}$). The evolution equation for the new invariants
can now be extracted by projecting the r.h.s. on the appropriate tensor
structure. We will not pursue this generalization in the present paper.

In appendix D we show that for $SU(2)$ this complication is
absent. There exists no additional invariant tensor ${\cal T}$, and one
finds the $n$-independent result
\be\label{C.4}
\sum^3_{l=1} \nu_l^{2m}=2\ee
for all $m=1,2...$.

\section*{Appendix D}

\renewcommand{\theequation}{D.\arabic{equation}}
\setcounter{equation}{0}

In this appendix we investigate in more detail the group-theoretical
quantities $\sum_l\nu_l^{2m}$ which occur in many calculations involving
covariantly
constant backgrounds of the type $A_\mu^z=n^z{\sf A}_\mu$.
Here we consider
an arbitrary (semi-simple, compact) gauge group $G$ with structure
constants
$f^{wyz}$. For a fixed unit vector $n^z$ we consider the matrix
\be\label{D.1}
\hat n^{yz}=n^w(T^w)^{yz}=if^{ywz}n^w\ee
The numbers $\nu_l,l=1,...,dim\ G$ are the eigenvalues of $\hat n$:
$\hat n^{yz}\psi^z_l=\nu_l\psi^y_l$. This equation can be rewritten in a
more
suggestive form. Let $t^z$ denote the generators of $G$ in an arbitrary
representation: $[t^w,t^y]=if^{wyz}t^z$. If we define
\be\label{D.2}
\tilde n=n^zt^z,\quad\tilde\psi_l=\psi^z_lt^z\ee
the eigenvalue equation becomes
\be\label{D.3}
[\tilde n,\tilde\psi_l]=\nu_l\tilde\psi_l\ee
Clearly the $\nu_l$'s do not depend on the representation chosen. We would
like to know how the spectrum $\lbrace\nu_l\rbrace$ depends on the
vector $n$. First of all, it is clear that if $V$ is any group element
in the $t$-representation, the matrices $\tilde n$ and $\tilde n'=
V\tilde nV^{-1}$ have
the same spectrum, i.e. the spectrum is constant along the orbit of $G$ in
the adjoint representation. If two directions $n$ and $n'$ are not
related by a group transformation, then the spectra
$\lbrace\nu_l(n)\rbrace$
and $\lbrace\nu_l(n')\rbrace$ can be different. Typically, for $G$ large
enough \cite{rai}, the orbit space is indeed nontrivial, and the spectrum
``feels'' the direction of $n^z$.

Let us go over from the basis $\lbrace T^z\rbrace$ to the Cartan-Weyl
basis $\lbrace H_i,E_{\vec\alpha}\rbrace$ of the abstract Lie algebra.
Here ${\vec\alpha}\in {\sf I\!R}^r$ are the root vectors and
$i=1,...,r\equiv\ rank\ G$. For definiteness we assume that the
$t^z$'s are in the fundamental representation where we write $\lbrace
h_i,e_{\vec\alpha}\rbrace$ for the Cartan-Weyl basis. Thus
\be\label{D.4}
[H_i,E_{\vec \alpha}]=\alpha_iE_{\vec\alpha}\quad{\rm and}\quad
[h_i,e_{\vec\alpha}]=\alpha_ie_{\vec\alpha}\ee
We assume that the generators $h_i$ of the Cartan subalgebra are given by
diagonal matrices. By an appropriate transformation $\tilde n\to V\tilde n
V^{-1}$ any $\tilde n$ can be brought to diagonal form. Therefore, in
order
to investigate  the $n$-dependence of $\lbrace\nu_l(n)\rbrace$, it is
sufficient to consider $\tilde n$'s which are in the Cartan subalgebra:
$\tilde n=\sum_{i=1}^rn_ih_i$. For this choice
\be\label{D.5}
[\tilde
n,e_{\vec\alpha}]=\left(\sum^r_{i=1}n_i\alpha_i\right)e_{\vec\alpha},
\quad[\tilde n,h_i]=0\ee
and the nonvanishing eigenvalues $\nu_l=\nu_{\vec\alpha}$ are given by
\be\label{D.6}
\nu_{\vec\alpha}=\sum^r_{i=1}n_i\alpha_i\ee
Therefore the quantities $\sum_l\nu_l^{2m}$ can be computed explicitly
from the root system:
\be\label{D.7}
\sum^{dim G}_{l=1}\nu^{2m}_l=\sum_{roots\{\vec\alpha\}}\left(
\sum^{rank\ G}_{i=1}n_i\alpha_i\right)^{2m}\ee

Let us consider a few simple examples. For $G=SU(2)$ we have
$r=1,n_1=1$ and there are only two (one-component) roots:
$\alpha=\pm1$. Thus
\be\label{D.8}
\sum^3_{l=1}\nu_l^{2m}=2,\quad m=1,2,3,... \ee
depends neither on the direction $n^z$ nor on the power $m$. This
degeneracy can be understood by noting that for $SU(2)$ the square of
the matrix (\ref{D.1}) is the projector $P_\perp$: $\hat n\hat n
=P_\perp$. This means that $\sum_l\nu^{2m}_l=\Tr(P^m_{\perp})=\Tr
(P_\perp)=2$, as it should be. Contracting eq. (\ref{C.3}) with
$n^{z_1}...n^{z_{2m}}$ and comparing the result to (\ref{D.8}) we see
that there exists no nontrivial invariant tensor ${\cal T}$ and
that $\tau_m=2$ for all $m$.

For $G=SU(3)$ we have $r=2$, and a 2-component unit vector $(n_1,n_2)$
specifies the direction of the field in the Cartan subalgebra. Using
the explicit form of the roots it is straightforward to derive that
\be\label{D.9}
\sum^8_{l=1}\nu^{2m}_l=2^{1-2m}\left[(n_1+\sqrt
3n_2)^{2m}+(n_1-\sqrt3n_2)^{2m}+(2n_1)^{2m}\right]\ee
For $m=1$ and $m=2$ it turns out that this expression depends on $n_1$
and $n_2$ only via $n^2_1+n^2_2=1$, and one obtains the
direction-independent
results
\be\label{D.10}
\sum^8_{l=1}\nu^2_l=3,\quad\sum^8_{l=1}\nu^4_l=\frac{9}{4}\ee
Starting from $m=3$, the invariants are explicitly $n$-dependent. Writing
$n_1=\cos\theta, n_2=\sin\theta$ we find for $m=3$
\ba\label{D.11}
\sum_{l=1}^8\nu_l^6=\frac{3}{16}\left[11\cos^6\theta+15\cos^4\theta\sin^2
\theta
+45\cos^2\theta\sin^4\theta+9\sin^6\theta\right]\ea

As discussed in Appendix C, the $n$-dependence is related to the
existence of a nontrivial invariant tensor ${\cal T}_{z_1...z_6}$.
However, we are not going to calculate the corresponding coefficient
$\tau_3$ here.  On the other  side we note that the definition of $\tau$
in (\ref{C.3}) depends on the specific truncation. Different truncations -
i.e. different choices of the definition of a term $\sim {\cal
T}_{z_1...z_6} F^{z_1}_{\mu\nu} F^{z_2}_{\mu\nu} F^{z_3}_{\rho\sigma}
F^{z_4}_{\rho\sigma}
F^{z_5}_{\tau\lambda}F^{z_6}_{\tau\lambda}$ whose coefficient is set to
zero - correspond to a different angle $\theta$ for  which the r.h.s. of
(\ref{D.11}) equals $\tau_3$. For $N=3$ we may therefore use
$\tau_2=\frac{9}{4}$ and $\tau_3$ between $\frac{27}{16}$ and
$\frac{33}{16}$, the last uncertainty reflecting the uncertainty from this
particular part of the truncation.

\section*{Appendix E}

\renewcommand{\theequation}{E.\arabic{equation}}
\setcounter{equation}{0}

In the regime where $\bar gB/k^2\grgl 1$ the use of the Euler-McLaurin
 expansion (\ref{3.9}) becomes questionable and we should look for an
alternative representation of the spectral sums (\ref{3.7}) and
(\ref{3.8}).
In this section we use the Schwinger proper-time representation
\cite{direu}.
It can be easily applied only for the simplified cutoff function
\be\label{E.1} R_k(x)=Z_kk^2.\ee
In this case one may write $(x\equiv {\cal D}_T)$
\be\label{E.2}
H(x)\equiv\frac{\partial_tR_k(x)}{x+R_k(x)}=\frac{\partial}{\partial
t}(Z_kk^2)\int^\infty_0dse^{-sZ_kk^2}e^{-sx}\ee
Inserting this representation into (\ref{3.7}) we may employ
\[\sum^\infty_{n=0}\exp[-sW_k'\bar g|\nu_l|B(2n+1)]=\frac{1}
{2\sinh[sW_k'\bar g|\nu_l|B]}\]
and
\[\int^\infty_0dxx^{\frac{d}{2}-2}e^{-sW_k'x}=v^{-1}_{d-2}2^{1-d}
\pi^{1-\frac{d}{2}}(sW_k')^{1-\frac{d}{2}}\]
as long as $W_k'>0$. One finds
\ba\label{E.3}
&&\Omega^{-1}\Tr_{xcL}[H(W_k'\D_T)]=2(4\pi)^{-\frac{d}{2}}(W_k')^{1
-\frac{d}{2}}\nonumber\\
&&\cdot\frac{\partial}{\partial t}(Z_kk^2)\sum^{N^2-1}_{l=1}\bar g|\nu_l|B
\int^\infty_0\frac{ds}{s}s^{(4-d)/2}e^{-sZ_kk^2}\nonumber\\
&&\cdot\left[\frac{d}{2\sinh(sW_k'\bar g|\nu_l|B)}-\exp(-sW_k'\bar
g|\nu_l|B)
+\exp(+sW_k'\bar g|\nu_l|B)\right]\ea
where the last exponential is due to the unstable
mode. For $Z_kk^2<W_k'\bar g
|\nu_l|B$ it makes the $s$-integration divergent at the upper (i.e. IR)
limit. In conventional calculations of the one-loop effective action
this creates a problem from the outset, because one attempts to work at
$k^2\to0$ there. In the present formulation everything is well defined
for $k^2$ sufficiently large, and one interesting question is how the
renormalization group flow behaves as one approaches $k^2\approx \theta$
from above.

In the UV limit $s\to 0$ the terms inside the square bracket
in (\ref{E.3}) behave as
\be\label{E.4}
[...]=\frac{d}{2\bar g|\nu_l|BW_k'(B^2/2)}\cdot\frac{1}{s}+O(s)\ee
While the $O(s)$-terms do not lead to UV-divergences for $d<6$, the
term $\sim 1/s$ leads to a divergent contribution to the proper-time
integral. Though the factor $\bar g|\nu_l|B$ cancels against a similar one
coming from the density of states, this divergent piece is still
field-dependent because of the $B$-dependence of $W_k'$.  This
UV divergence shows a failure of the truncation for
the mass-type cutoff function $R_k=Z_kk^2$. The latter may
be used only together with the approximation $W_k''=0$
on the r.h.s. of the flow equation. The divergent
piece in the proper-time integral is an irrelevant constant then.

Using (\ref{E.3}) and a similar formula for the scalar traces in eq.
(\ref{3.4}), we obtain (up to an irrelevant constant and for $\tilde Z
_k=1/\alpha_k,\tilde\eta=0$)
\ba\label{E.5}
&&\frac{\partial}{\partial t}W_k(\frac{1}{2}B^2)=(4\pi)^{-d/2}\sum^{N^2-1}
_{l=1}\bar g|\nu_l|Bk^2\int^\infty_0ds\ s^{1-d/2}\cdot\nonumber\\
&&\cdot\left\lbrace(2-\eta)Z_k(W_k')^{1-\frac{d}{2}}\left[\frac{d-1}
{2\sinh(sW_k'\bar g|\nu_l|B)}+2\sinh(sW_k'\bar g|\nu_l|B)\right]
e^{-sZ_kk^2}\right.
\nonumber\\
&&\left.-\frac{\exp(-sk^2)}{\sinh(s\bar g
|\nu_l|B)}\right\rbrace\ea
This evolution equation is the analogue of (\ref{3.11}) with the
additional
assumption $W_k''=0$. Contrary to the Euler-McLaurin series it is valid
even
for strong fields $\bar gB\approx k^2$.

For $\bar gB\ll k^2$ the r.h.s. of (\ref{E.5}) can be expanded in powers
of $B$. Apart from the different form of $R_k$, this reproduces the
Euler-McLaurin
expansion. Expanding up to order $B^4$ we find, for instance,
\be\label{E.6}
\frac{\partial}{\partial t}\left(\frac{w_2}{g^2}\right)=4\frac{w_2}{g^2}
+\frac{127}{360\pi^2}\tau_2r_0^{4,2}\ee
with $r_0^{4,2}=2$.
This result is the counterpart of eq. (\ref{4.11}) which had been obtained
with the exponential cutoff for which $r_0^{4,2}=1/6$. In accordance with
(\ref{3.13}) we find an additional factor of 12 in the second term on the
r.h.s. of (\ref{E.6}). Hence also the value of the fixed point $w_{2*}(k)$
is 12 times larger than the result (\ref{4.12}). In view of the discussion
following (\ref{4.13}) this means that for the mass-type cutoff
$R_k=Z_kk^2$
this higher-order correction is much larger than for the exponentially
decreasing $R_k$ of (\ref{1.2}). This is probably closely related to
the ultraviolet problems and
indicates that, though computationally
more difficult to handle, the exponential
cutoff (\ref{1.2}) should be used for reliable estimates.

\section*{Appendix F}

\renewcommand{\theequation}{F.\arabic{equation}}
\setcounter{equation}{0}

In this appendix we derive the flow equation in the $F^6$ truncation.
We start from the evolution equation for $\ddot w(\vartheta)$ which
follows from differentiating (\ref{4.7}) with respect to $\vartheta$:
\ba\label{F.1}
&&\frac{\partial}{\partial t}\ddot w=(4+\eta)\ddot w+4\vartheta w^{(3)}
\nonumber\\
&&-(2-\eta)v_dg^2\dot w^{-\frac{d}{2}}\left\lbrace(d-2)\sum^\infty_{m=1}
\tau_m(C^d_m-E_m)r^{d,m}_0\right.\nonumber\\
&&\left[(2\vartheta\dot w^2)^{m-1}\left(((8-d)m-d)\dot w\ddot
w+((4-2d)m-d+
\frac{d^2}{2})\vartheta\ddot w^2+(4m-d)\vartheta\dot w
w^{(3)}\right)\right.
\nonumber\\
&&+2(m-1)(2\vartheta\dot w^2)^{m-2}\dot w^2(\dot w+2\vartheta\ddot
w)(2m\dot w+(4m-d)\vartheta\ddot w)\Biggr]\nonumber\\
&&+r^d_1\left[\frac{4w^{(3)}+2\vartheta w^{(4)}}{\dot w+2\vartheta \ddot
w}-
\frac{4(\ddot w+\vartheta w^{(3)})(3\ddot w+2\vartheta
w^{(3)})+2\vartheta\ddot w(5w^{(3)}+2\vartheta w^{(4)})}{(\dot
w+2\vartheta\ddot w)^2}\right.\nonumber\\
&&+\frac{4\vartheta\ddot w(3\ddot w+2\vartheta w^{(3)})^2}{(\dot
w+2\vartheta
\ddot w)^3}\nonumber\\
&&\left.-d\frac{2\ddot w^2+3\vartheta\ddot w w^{(3)}}{\dot w(\dot
w+2\vartheta
\ddot w)}+\frac{3}{2}d\frac{\vartheta\ddot w^3}{\dot w^2(\dot
w+2\vartheta\ddot w)}+2d\frac{\vartheta\ddot w^2(3\ddot w+2\vartheta
w^{(3)})}{\dot w(\dot w+2\vartheta\ddot w)^2}\right]\nonumber\\
&&\left.+\frac{1}{4}d(d-1)(d-2)(N^2-1)r^d_2\left[\frac{w^{(3)}}{\dot w}-
\frac{d+2}{2}\frac{\ddot w^2}{\dot w^2}\right]\right\rbrace\nonumber\\
&&+8(d-2)v_dg^2\sum^\infty_{m=1}m(m-1)
\tau_mE_mr^{d,m}_0(2\vartheta)^{m-2}\ea
Evaluating this equation for $\theta=0$ yields eq. (\ref{5.8}).
Taking one further $\vartheta$-derivative at $\vartheta=0$ we obtain
for $d=4$ and $k>k_{np}$ the flow equation for $w_3$ (with $C^4_2-E_2=
-\frac{29}{80},\ C^4_3-E_3=-\frac{137}{10080}$
and $E_3=\frac{31}{30240})$:
\ba\label{F.2}
&&\frac{\partial}{\partial t}w_3=(8+\eta)w_3\nonumber\\
&&+\frac{g^2}{16\pi^2}\left\lbrace\left(\frac{442}{315}-\frac{137}{210}
\eta\right)\tau_3r^{4,3}_0+\frac{87}{5}(2-\eta)\tau_2r^{4,2}_0w_2
\right.\nonumber\\
&&-(2-\eta)r^4_1(3w_4-51w_2w_3+105w^3_2)\nonumber\\
&&-3(2-\eta)(N^2-1)r^4_2(w_4-9w_2w_3+12w^3_2)\Biggr\rbrace\ea
For the infrared cutoff (\ref{1.2}) one has $r^{4,2}_0=\frac{1}{6},
r_0^{4,3}
=-\frac{1}{30}$ and we observe that the perturbatively leading term
$\sim g^2$ is negative. This implies a positive perturbative fixpoint
value
for the ratio $w_3/g^2$.

This discussion can easily be generalized for
arbitrary $w_n$. In lowest order in $g^2$ the flow equation (\ref{F.1})
simplifies considerably $(k>k_{np})$
\ba\label{F.3}
&&\frac{\partial}{\partial t}\ddot w=4\ddot w+4\vartheta
w^{(3)}\nonumber\\
&&-8(d-2)v_dg^2\sum^\infty_{m=2}m(m-1)\tau_mr^{d,m}_0
\left(C^d_m-2E_m\right)
(2\vartheta)^{m-2}\ea
This implies for $n\geq 2$ the flow equations
\be\label{F.4}
\frac{\partial}{\partial t}w_n=4(n-1)w_n-(d-2)v_dg^22^{n+1}n!\tau_n
r^{d,n}_0\left(C^d_n-2E_n\right)\ee
and the infrared fixed point values
\be\label{F.5}
\left(\frac{w_n}{(d-2)v_dg^2}\right)_*=
\frac{2^{n-1}}{n-1}n!\tau_nr^{d,n}_0\left(C^d_n
-2E_n\right)\ee
For $d=4$ and with (\ref{3.14}) this yields
\ba\label{F.6}
w_{n*}&=&d_n\frac{g^2}{16\pi^2}\nonumber\\
d_n&=&\frac{2^n\ n!}{(n-1)(2n-1)!}\tau_nB_{2n-2}\left(\frac{2^{2n-1}-1}
{2n}B_{2n}-1\right)\ea
where for $SU(2)$
\be\label{F.7}
\frac{d_2}{\tau_2}=-\frac{127}{540},
\quad\frac{d_3}{\tau_3}=\frac{221}{37800}
\ee
The series of $d_n$ is alternating as long as the bracket is dominated
by -1.

The infrared fixed point in $w_2/g^2$ implies that the coefficient $W_2$ in
(\ref{4.1}) diverges $\sim k^{-4}$
\be\label{4.14}
W_2\sim -\frac{\bar g^4}{k^4}\ee
and similar for $W_3$.
Because of the infrared divergence for $k\to 0$, a result of this type
could never have been found in standard perturbation theory. It could,
however, be derived using the
``$\frac{D}{Dt}\approx\frac{\partial}{\partial t}$''-approximation of the
effective average action which we displayed
in eq. (\ref{1.4}). For this purpose one can neglect the $W_k''$ terms on
the r.h.s. of (\ref{1.4}) and approximate $W_k'=w_1$. Then, with
(\ref{2.3})
inserted into (\ref{1.4}), one obtains for the $k$-dependent terms in
$\Gamma_k$:
\ba\label{4.15}
\Gamma_k[A,A]&=&\frac{1}{2}{\Tr}_{xcL}\ln\left[w_1{\D}_T+(w_1-
\frac{1}{\alpha_k})D\otimes
D+R_k(w_1{\D}_T+(w_1-\frac{1}{\alpha_k})D\otimes D)\right]\nonumber\\
&&-{\Tr}_{xc}\ln\left[-D^2+R_k(-D^2)\right]\ea
If one extracts the $F^4_{\mu\nu}$-term from these traces, one finds
a (renormalized) coefficient which equals exactly $W_2$,
as extracted from the fixed point (\ref{4.12}). Clearly
(\ref{4.10}) goes beyond the
$``\frac{D}{Dt}\approx\frac{\partial}{\partial t}''$ approximation. The
terms proportional to $\eta$ and to $w^2_2$ could
not have been obtained in this approximation.

Finally, for $k<k_{np}$ the flow equation for $w_3$ follows from
(\ref{F.1})
as
\ba\label{F.8}
&&\frac{\partial}{\partial t}w_3=8w_3+\frac{g^2}{16\pi^2w^2_1}\left\lbrace
\tau_3r_0^{4,3}\left(\frac{137}{105}w_1^6+\frac{31}{315}w^2_1\right)
+\frac{174}{5}\tau_2r^{4,2}_0w^3_1w_2\right.\nonumber\\
&&-6r_1^4\left(\frac{w_4}{w_1}-17\frac{w_2w_3}{w^2_1}
+35\frac{w^3_2}{w_1^3}\right)\nonumber\\
&&\left.-6(N^2-1)r^4_2\left(\frac{w_4}{w_1}-9\frac{w_2w_3}
{w_1^2}+12\frac{w^3_2}{w_1^3}\right)\right\rbrace\ea

\end{document}